\documentclass[12pt]{article}

\usepackage{amsmath,amssymb,graphicx} %use drftcite in draftmode otherwise
\usepackage{epsf}
\usepackage{cite}

\newcommand{\beq}{\begin{eqnarray}}% can be used as {equation} or {eqnarray}
\newcommand{\eeq}{\end{eqnarray}}

\newcommand{\centeron}[2]{{\setbox0=\hbox{#1}\setbox1=\hbox{#2}\ifdim
                           \wd1>\wd0\kern.5\wd1\kern-.5\wd0\fi \copy0
                           \kern-.5\wd0\kern-.5\wd1\copy1\ifdim\wd0>\wd1
                           \kern.5\wd0\kern-.5\wd1\fi}}
\newcommand{\ltap}{\>\centeron{\raise.35ex\hbox{$<$}}
                   {\lower.65ex\hbox{$\sim$}}\>}
\newcommand{\gtap}{\>\centeron{\raise.35ex\hbox{$>$}}
                   {\lower.65ex\hbox{$\sim$}}\>}

\newcommand{\lsim}{\mathrel{\ltap}}

\newcommand\ZZ{\hbox{\zfont Z\kern-.4emZ}}
\font\zfont = cmss10 %scaled \magstep1

\begin{document}
\begin{titlepage}
\begin{flushright}
{\tt hep-ph/0303236} \\
\end{flushright}

\vspace*{0.2cm}
\begin{center}
\vspace*{0.5cm}
{\LARGE \bf Variations of Little Higgs Models and their} \\
\vspace*{0.2cm}
{\LARGE \bf Electroweak Constraints} \\
\vspace*{1.5cm}

\mbox{\bf {Csaba Cs\'aki}$^{a}$, {Jay Hubisz}$^{a}$,
{Graham D. Kribs}$^{b}$,}\\
\mbox{\bf {Patrick Meade}$^{a}$, {and John Terning}$^{c}$} \\

\vspace*{0.8cm}

$^{a}$ {\it Newman Laboratory of Elementary Particle Physics, \\
Cornell University, Ithaca, NY 14853} \\
\vspace*{0.1cm}
$^{b}$ {\it Department of Physics, University of Wisconsin, Madison, WI 53706} \\
\vspace*{0.1cm} $^{c}$ {\it Theory Division T-8, Los Alamos
National Laboratory, Los Alamos,
NM 87545} \\
\vspace*{0.8cm} {\tt  csaki@mail.lns.cornell.edu,
hubisz@mail.lns.cornell.edu, kribs@physics.wisc.edu,
meade@mail.lns.cornell.edu, terning@lanl.gov}
\end{center}
\vspace*{.1cm}
\begin{abstract}
\vskip 3pt \noindent

We calculate the tree-level electroweak precision constraints on a
wide class of little Higgs models including: variations of the
Littlest Higgs $SU(5)/SO(5)$, $SU(6)/Sp(6)$, and
$SU(4)^4/SU(3)^4$. By performing a global fit to the precision
data we find that for generic regions of the parameter space the
bound on the symmetry breaking scale $f$ is several TeV, where we
have kept the normalization of $f$ constant in the different
models. For example, the ``minimal'' implementation of
$SU(6)/Sp(6)$ is bounded by $f>3.0$ TeV throughout most of the
parameter space, and  $SU(4)^4/SU(3)^4$ is bounded by $f^2 \equiv
f_1^2+f_2^2 > (4.2 \; \mathrm{TeV})^2$.  
In certain models, such as $SU(4)^4/SU(3)^4$,
a large $f$ does not directly imply a large
amount of fine tuning since the heavy fermion masses that
contribute to the Higgs mass can be lowered below $f$ for a
carefully chosen set of parameters. We also find that for certain
models (or variations) there exist regions of parameter space in
which the bound on $f$ can be lowered into the range $1$-$2$ TeV\@.
These regions are typically characterized by a small mixing
between heavy and standard model gauge bosons, and a small (or
vanishing) coupling between heavy $U(1)$ gauge bosons and the
light fermions. Whether such a region of parameter space is
natural or not is ultimately contingent on the UV completion.

\end{abstract}
\end{titlepage}
\newpage
%\renewcommand{\thefootnote}{(\arabic{footnote})}

%%%%%%%%%%%%%%%%%%%%%%%%%%%%%%%%%%%%%%%%%%%%%%%%%%%%%%
%%%%%%%%%%%%%%%%%%%%%%%%%%%%%%%%%%%%%%%%%%%%%%%%%%%%%%
\section{Introduction}
\label{sec:intro}
\setcounter{equation}{0}
 Hierarchies in masses are ubiquitous in the Standard Model (SM). Fortunately, symmetries
prevent gauge bosons and matter fermions from acquiring radiative
corrections to their masses beyond logarithmic sensitivity to
heavy physics.  The Higgs boson mass, however, is quadratically
sensitive to heavy physics.  Hence, naturalness suggests the
cutoff scale of the SM should be only a loop factor
higher than the Higgs mass.  However, there are many probes of
physics beyond the SM at scales ranging from a few to
tens of TeV\@. In particular, four-fermion operators that give rise
to new electroweak (EW) contributions generally constrain the new
physics scale to be more than a few TeV, and some new flavor-changing
four-fermion operators are constrained even further, to be above
the tens of TeV level.  With mounting  evidence for the
existence of a light Higgs $\lsim 200$ GeV \cite{LEPEWG}, we are
faced with understanding why the Higgs mass is so light compared
with radiative corrections from cutoff-scale physics that appears
to have been experimentally forced to be above the tens of TeV
level.  The simplest solution to this ``little hierarchy problem''
is to fine-tune the bare mass against the radiative
corrections, but is widely seen as being unnatural.

There has recently been much interest
[2-15]
%\cite{little1,other1,minmoose,littlest,other2,
%su6sp6,dudes,wackchang,bigcorrections,%
%hewett,burdman,wisconsin,Dib:2003zj,wisc2}
in a new approach to solving the little hierarchy problem, called
little Higgs models.  These models have a larger gauge group
structure appearing near the TeV scale into which the EW
gauge group is embedded.  The novel feature of little Higgs models
is that there are approximate global symmetries that protect the
Higgs mass from acquiring \emph{one-loop} quadratic sensitivity to
the cutoff.  This happens because the approximate global
symmetries ensure that the Higgs can acquire mass only through
``collective breaking'', or multiple interactions.  In the limit
that any single coupling goes to zero, the Higgs becomes an exact
(massless) Goldstone boson. Quadratically divergent contributions
are therefore postponed to two-loop order, thereby relaxing the
tension between a light Higgs mass and a cutoff of order tens of
TeV\@.

\begin{figure}
\centerline{\includegraphics[width=1.0\hsize]{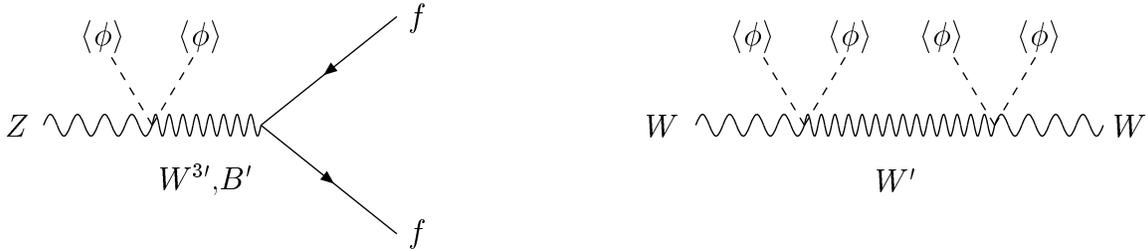}}
\label{naive-fig} \caption{Corrections to $\Gamma_Z$ and to
$M_W^2$ through mixing with heavy gauge bosons.}
\end{figure}

The minimal ingredients of little Higgs models appear to be additional
gauge bosons, vector-like colored fermions, and additional
Higgs doublets and/or Higgs triplets, as well as scalars uncharged
under the SM gauge group.
In general modifications of the EW sector are usually tightly constrained
by precision EW data (see Refs.~[16-19]
%\cite{Sekhar,CST,RSfit,CEKT}
for example).
One generic feature, new heavy gauge
bosons, can be problematic if the SM gauge bosons mix with them or
if the SM fermions couple to them. This is easy to see:  Consider
the modification to the coupling of a $Z$ to two fermions and
(separately) the modification to the vacuum polarization of the $W$,
as shown in Fig.~\ref{naive-fig}.  These are among the best
measured EW parameters that agree very well with the
SM predictions (using $M_Z$, $G_F$, and $\alpha_{em}$
as inputs):  both of these observables have been measured to
$\pm0.2\%$ to 95\% C.L.  Generically the corrections to these
observables due to heavy $U(1)$ gauge bosons and heavy $SU(2)$ gauge
bosons can be simply read off from Fig.~\ref{naive-fig} as
\begin{equation}
\frac{\delta \Gamma_Z}{\Gamma_Z} \sim 1 + c_1 \frac{v^2}{f^2}
\quad , \quad \frac{\delta M_W^2}{M_W^2} \sim 1 + c_2
\frac{v^2}{f^2} \; ,
\end{equation}
where $f$ is roughly the mass of the heavy gauge boson, and $c_1$ and
$c_2$ parameterize the strength of the couplings between
heavy-to-light fields.  For $c_1 \sim 1$ or $c_2 \sim 1$, it is
trivial to calculate the EW bound on $f$,
\begin{equation}
f > 5.5 \; \mbox{TeV} \quad \mbox{to 95\% C.L.} \; .
\end{equation}
Notice that even if the coupling of light fermions to the heavy
gauge bosons were zero ($c_1 = 0$), maximal mixing among $SU(2)$
gauge bosons ($c_2 = 1$) is sufficient to place a strong
constraint on the scale of new physics.

In our previous paper \cite{bigcorrections} we examined the
tree-level precision EW constraints on the Littlest Higgs
model, $SU(5)/SO(5)$.  We found strong constraints on the symmetry
breaking scale $f$ consistent with the above naive argument.  The
reason for the appearance of these large corrections is that some
interactions involving the heavy gauge bosons violate custodial
$SU(2)$. However, the custodial
$SU(2)$ violating corrections come mostly from the exchange of the
heavy $U(1)$ gauge boson,
thus one might try to adjust the $U(1)$ sector of the theory so that the
contributions to the EW precision observables can be
reduced. We examine such possibilities for the modifications of
the $U(1)$ sector in the Littlest Higgs model in the first part of
the paper, by including other $U(1)$ fermion charge assignments,
gauging only $U(1)_Y$, and gauging a different combination of
$U(1)$'s.  In the second part of the paper we consider changing
the global symmetry structure slightly (to the $SU(6)/Sp(6)$
model) and more drastically (the $SU(4)^4/SU(3)^4$ ``simple
group'' model). Generically, meaning no special choices of the
model parameters, we find that these models have constraints
comparable to the those on the Littlest Higgs model. Unlike our
previous analysis, however, we find regions of parameter space for
certain models (or their variations) in which the bound on the
symmetry breaking scale $f$ is lowered to $1$-$2$ TeV\@.  We identify
the extent of these regions of parameter space.  Most recently a
model based on $SO(5)^8/SO(5)^4$ which has a custodial $SU(2)$ has
been proposed~\cite{wackchang}.  This model was specifically
constructed to avoid constraints on the model from custodial
$SU(2)$ violation from heavy gauge boson exchange.  However, this
model does contain $SU(2)$ triplets that could in principle lead
to constraints, a complete analysis will be given
in~\cite{future}.

The organization of this paper is as follows.  In
Section~\ref{u1little} we review the Littlest Higgs model and
consider the various possible $U(1)$ charge assignments. In
Section~\ref{modlittlest} we examine what happens when different
choices are made for the $U(1)$ gauge structure,
including the case that the only gauged $U(1)$  is the standard
hypercharge. In Section~\ref{su6sp6sec} we study the $SU(6)/Sp(6)$
model which has two Higgs doublets but no Higgs triplet. In
Section~\ref{dudes-sec} we consider the $SU(4)^4/SU(3)^4$ model.
Finally, in Section~\ref{conclusions-sec} we conclude and discuss
the implications of our results.

%%%%%%%%%%%%%%%%%%%%%%%%%%%%%%%%%%%%%%%%%%%%%%%%%%%%%%
%%%%%%%%%%%%%%%%%%%%%%%%%%%%%%%%%%%%%%%%%%%%%%%%%%%%%%

%%%%%%%%%%%%%%%%%%%%%%%%%%%%%%%%%%%%%%%%%%%%%%%%%%%%%%
%%%%%%%%%%%%%%%%%%%%%%%%%%%%%%%%%%%%%%%%%%%%%%%%%%%%%%
\section{The Littlest Higgs: varying the U(1) embedding of the
SM fermions}\label{u1little} \setcounter{equation}{0}
\par
In the Littlest Higgs model, large contributions to EW
precision observables arise from the exchange of the heavy $U(1)$
gauge boson \cite{bigcorrections,hewett}.  These large corrections
are due to the custodial $SU(2)$ violating effects of the broken
gauge sectors.  Custodial
$SU(2)$ violations form the
heavy $SU(2)$ sector appear only at order $v^4/f^4$, which is
negligible compared to the leading corrections of order
$v^2/f^2$. These leading corrections arise from the exchange of the
heavy $U(1)$ gauge boson,  thus the $U(1)$ sector of
the theory is the most problematic to EW precision
constraints. The first modification we consider is to change the
$U(1) \times U(1)$ charge assignments of the SM fermions. This has
the potential to relax the bounds from EW precision
observables by reducing the effective coupling to the heavy $U(1)$
gauge boson. In this section we examine this possibility after
briefly reviewing the structure of the Littlest Higgs model.

\subsection{The Littlest Higgs model}

The Littlest Higgs model is based on the non-linear $\sigma$ model
describing an $SU(5)/SO(5)$ global symmetry
breaking~\cite{littlest}. This symmetry breaking can be thought of
as originating from a VEV of a symmetric tensor of the $SU(5)$
global symmetry. A convenient basis for this breaking is
characterized by the direction $\Sigma_0$ given by
\begin{equation}
\Sigma_0 =\left( \begin{array}{ccccc} &&&1& \\&&&&1\\ &&1\\1\\&1
\end{array} \right).
\end{equation}
The Goldstone fluctuations are then described by the pion fields
$\Pi = \pi^a X^a$, where the $X^a$ are the broken generators of
$SU(5)$. The non-linear sigma model field is then
\begin{equation}
\Sigma (x) = e^{i\Pi/f} \Sigma_0 e^{i \Pi^T/f}=e^{2i\Pi/f}
\Sigma_0.
\end{equation}
where $f$ is the scale of the VEV that accomplishes the breaking.
An $[SU(2)\times U(1)]^2$ subgroup of the $SU(5)$ global symmetry
is gauged, where the generators of the gauged symmetries are given
by
\begin{eqnarray}
&Q_1^a=\left( \begin{array}{ccc} \sigma^a/2 &0 & 0 \\
0 & 0 & 0\\ 0 & 0 & 0
\end{array}\right), \ \ \ &Y_1=
{\rm diag}(-3,-3,2,2,2)/10\nonumber \\
&Q_2^a=\left( \begin{array}{ccc} 0 & 0 & 0\\
0 & 0 & 0 \\
0 &0&-\sigma^{a*}/2\end{array} \right), & Y_2={\rm
diag}(-2,-2,-2,3,3)/10~,
\end{eqnarray}
where $\sigma^a$ are the Pauli $\sigma$ matrices. The $Q^a$'s are
$5 \times 5$ matrices written in terms of $2 \times 2$, 1, and $2
\times 2$ blocks. The Goldstone boson matrix $\Pi$, in terms of
the uneaten fields, is then given by
\begin{equation}
\Pi = \left( \begin{array}{ccc} 0 & \frac{H^\dagger}{\sqrt{2}} &
\phi^\dagger
\\ \frac{H}{\sqrt{2}}& 0 & \frac{H^*}{\sqrt{2}}\\ \phi
&\frac{H^T}{\sqrt{2}} & 0
\end{array}\right),
\end{equation}
where $H$ is the little Higgs doublet $(h^0,h^+)$ and $\phi$ is a
complex triplet Higgs, forming a symmetric tensor $\phi_{ij}$.

The kinetic energy term of the non-linear $\sigma$ model is \beq
\frac{f^2}{8} {\rm Tr} D_\mu \Sigma (D^\mu \Sigma)^\dagger \eeq
where \beq D_\mu \Sigma = \partial_\mu \Sigma - i \sum_j \left[
g_j W_j^a (Q_j^a \Sigma + \Sigma Q_j^{aT} ) + g_j^\prime B_j( Y_j
\Sigma + \Sigma Y_j)\right]~, \eeq and $g_j$ and $g_j'$ are the
couplings of $[SU(2)\times U(1)]_j$ gauge groups.

This structure of the gauge sector prevents quadratic divergences
arising from gauge loops. In order to cancel the divergences due
to the top quark the top Yukawa coupling is obtained from the
operator
\begin{equation}\label{topyuk}
{\cal L}_{top} =\frac{1}{2}\lambda_1 f \epsilon_{ijk}\epsilon_{xy}
\chi_i \Sigma_{jx} \Sigma_{ky} u_3'^c +\lambda_2 f
\tilde{t}\tilde{t}^c + \mbox{h.c.},
\end{equation}
where $\chi=(b_3\;t_3\;\tilde{t})$, that preserves enough of the
global symmetry to forbid a one-loop quadratic divergence arising from
the top quark. The potential that gives rise to the Higgs quartic
scalar interaction comes from the quadratically divergent terms
in the Coleman-Weinberg (CW) potential and their tree level
counterterms. Evaluating the contributions from both the gauge and
fermion sector we find that the gauge loops contribute
\begin{equation}
a f^2\left[ (g_2^2+g_2'^2) \left| \phi_{ij}+\frac{i}{4f}( h_i h_j+
h_j h_i)\right|^2+ (g_1^2+g_1'^2) \left| \phi_{ij}-\frac{i}{4f}(
h_i h_j+ h_j h_i)\right|^2\right], \label{CW1}
\end{equation}
where $a$ is an order one constant determined by the relative size
of the tree-level and loop-induced terms.  Similarly, the fermion
loops contribute
\begin{equation}
-a' \lambda_1^2 f^2\left| \phi_{ij}+\frac{i}{4f}( h_i h_j+ h_j
h_i)\right|^2, \label{CW2}
\end{equation}
where $\lambda_1$ (and $\lambda_2$) are the Yukawa couplings and
mass terms.  The $SU(3)$ global symmetries enforce the two
possible combinations of $h$ and $\phi$ in the potential given
above. Therefore two parameters $a$ and $a'$ are sufficient to
completely parameterize the potential.

\subsection{Fermion U(1) charges}

The $U(1)$ charges of the SM fermions are constrained by requiring that
the Yukawa couplings are gauge invariant and maintaining the usual
SM hypercharge assignment.  The latter imposes the constraint
$y_1+y_2=Y_{SM}$.  For the top quark, the Yukawa coupling is fixed
by the global symmetries (\ref{topyuk}), and hence its $U(1)$
charges are fixed.  Furthermore, if mixed SM gauge group/$U(1)_i$
anomalies are to be avoided, the entire third generation $U(1)$
charges can be determined. We find that the $U(1)\times U(1)$ charges
of the third generation are $Q:(\frac{1}{10},\frac{1}{15})$,
$u_3:(-\frac{2}{5},-\frac{4}{15})$. The Yukawa couplings for the
first and second generations and the down and lepton sectors can
be written identically as in (\ref{topyuk}) with only the change
of $\Sigma \to \Sigma^*$ in the down and lepton sectors and an
extra fermion introduced for all the SM particles to cancel the
one loop quadratic divergences. The charge assignments are
determined just as they are for the third generation.

However, the quadratic divergences for the first two generations
are much smaller than that of the top quark due to the small size
of their Yukawa couplings.  One could therefore ignore these
numerically irrelevant quadratic divergences.  This means the
Yukawa couplings of the first two generations of fermions do not
need to respect the global symmetries, and thus need not be of the
form (\ref{topyuk}).  Hence, the $U(1)$ charges of these fields
could be modified~\cite{Nima}.

Here we consider modifying the $U(1)$ charges of the first two
generations to be different from the third generation.  The same constraints
enter as before, namely that the Yukawa couplings are gauge invariant
and each light fermion retains its usual SM hypercharge as it must.
The charges of a light fermion
$F$ can be written as $R Y_F$ under the first $U(1)$ and $(1-R)
Y_F$ under the second $U(1)$, where $Y_F$ is the SM hypercharge of
the fermion.  We will also assume that $R$ is universal within
each generation of fermions. This is the simplest assignment that
avoids mixed SM gauge group/$U(1)_i$ anomalies.  Also, we do not
consider a different $R$ between the first two generations
since this would lead to new contributions to flavor changing
neutral current processes.

The possible values of $R$ can be determined by requiring the
invariance of the Yukawa couplings under the $U(1)$'s. Given our
assumptions above, it is sufficient to consider one light fermion
Yukawa coupling, such as for up-type quarks
\begin{equation}\label{miniyuk}
q h X^r Y^s u^c
\end{equation}
where $X$ and $Y$ are the components of the $\Sigma$ field that
get VEVs of order $f$, while $r$ and $s$ are assumed to be
integers.  Here, the field $X$ corresponds to the $(1,4)$ and
$(2,5)$ components of the $\Sigma$ field and $Y$ corresponds to
the $(3,3)$ component. The $U(1)$ charges of $X$ are
$(\frac{1}{10},-\frac{1}{10})$ and for $Y$ they are
$(\frac{4}{10},-\frac{4}{10})$. The $U(1)$ charges of the Higgs
can be read off from its embedding into $\Sigma$ as either
$(\frac{4}{10},\frac{1}{10})$ or $(\frac{1}{10},\frac{4}{10})$.
(The two possibilities for the Higgs charge assignment are present
because the light Higgs is a mixture of two fields with different
$U(1) \times U(1)$ charge after the two $U(1)$'s are broken to the
diagonal subgroup.) Assuming the Higgs field in
(\ref{miniyuk}) has the $U(1)$ charges
$(\frac{4}{10},\frac{1}{10})$ (as would be the case for an
operator of the type (\ref{topyuk}) with additional powers of $X$
and $Y$), we find that
\begin{equation}\label{realrule}
R=(4-r+4s)/5,
\end{equation}
i.e., $R$ can only take on integer multiples of $1/5$.  Similarly,
an operator involving the Higgs field with the other $U(1) \times
U(1)$ charge assignment gives
\begin{equation}
R=(1-r+4s)/5 .
\end{equation}
In either case, only integer multiples of $1/5$ for $R$ are
allowed. For the third generation, the top Yukawa coupling
(\ref{topyuk}) corresponds to (\ref{miniyuk}) with $r=1$ and
$s=0$, and thus $R=3/5$. Therefore $R=3/5$ is the only value of
$R$ whereby quadratic divergences can be canceled for fermions of
every generation.  Nevertheless we stress that this result is a
consequence of our assumption that the modification of the $U(1)$
charges is universal within a generation. (If we drop this
requirement there will be no simple constraints on the values of
$R$.)

In~\cite{bigcorrections} we calculated the tree-level EW
precision constraints on the Littlest Higgs model assuming that
the first two generation of fermions transformed only under the
first gauge group, hence $R=1$.  For this choice of light fermion
$U(1)$ charges we found the bound on $f$ to be 4 TeV with
a fine-tuning in the Higgs mass of less than a percent.  We now
redo our calculations for the corrections to EW precision
observables for generic values of $R$.  As we explained in
\cite{bigcorrections}, the main quantities that are necessary to
compute the EW precision observables are: the $W$ and $Z$
masses $M_W$ and $M_Z$, the Fermi coupling $G_F$, the shifts in
the couplings of the $Z$ boson to the light fermions $\delta
\tilde{g}^{FF}$, and the low-energy neutral current Lagrangian
parameterized in terms of $\rho_*$ and $s_*^2(0)$. The detailed
definition of these quantities, as well as the observed and bare
values of the weak mixing angle $s_0^2$ and $s_W^2$ is given in
[10,22-26].
%\cite{bigcorrections,Sformulas,ErlerLang,Lynn,Burgess,GAPP}.
We find the above quantities re-expressed as a
function of $R$ to be:
\beq M_W^2 &=& \frac{g^2
v^2}{4}\left(1+\frac{\Delta}{4}(1+4c^2-4c^4)+
4\Delta'\right)\nonumber \\
M_Z^2 &=&
\frac{(g^2+g'^2)v^2}{4}\left(1+\frac{\Delta}{4}(1+4c^2-4c^4)-
\frac{5 \Delta}{4}(1-2c'^2)^2+8\Delta'\right)\nonumber \\
\delta \tilde{g}^{FF} &=& \Delta/2 \left[ (5 (c'^2 s'^2+c'^4(R-1)
-s'^4 R)+ c^2(1-2c^2))T_3^F \right. \nonumber \\ &&
\,\,\,\,\,\,\,\,\,\,\,\,\,\,\,\,  \left. -5 (c'^2 s'^2+c'^4(R-1)
-s'^4 R)Q_F
\right] \nonumber
\eeq
\beq
\rho_*&=&1-4\Delta'+\frac{5}{4} \Delta (1-2R)^2 \nonumber \\
s_*^2 (0)&= &s_W^2+\frac{\Delta}{2} \left[ 5R (s_W^2-1) (1-2R)+ 5
c'^2(s_W^2-1)(2R-1)-c^2 s_W^2\right]\nonumber \\
s_0^2&=&s_W^2-\frac{s_W^2c_W^2}{c_W^2-s_W^2}\left[\Delta\left(c^2-c^4-\frac{5}{4}(1-2c'^2)^2\right)+4\Delta'\right]\nonumber \\
G_F&=&\frac{1}{\sqrt{2}v^2}\left(1-\frac{\Delta}{4}-4\Delta'\right)
~. \label{U1corrections}
\eeq
where we have defined (as in \cite{bigcorrections})
\begin{equation}
\Delta = \frac{v^2}{f^2}; \;\; \Delta'=\frac{v'^2}{v^2}; \;\;
c=\frac{g}{g_2}; \;\; c'=\frac{g'}{g_2'},
\end{equation}
and $v^\prime$ is the VEV of the scalar $SU(2)$ triplet.  Using these
expressions all the EW precision observables can be obtained as in
\cite{bigcorrections}.
\par
The largest contribution to the constant shift in the observables
listed in the Appendix of~\cite{bigcorrections} comes from the
$U(1)$ sector of the theory (plus the constant shift coming from
the triplet VEV).  In the following section we calculate the
bounds on $f$ for a general $R$.  The light fermion coupling to
the heavy $B$ vanishes for $R=1/2$ and $c^\prime=1/\sqrt{2}$.  The
decoupling of the corrections resulting from exchange of the heavy
$B$ gauge boson is evident in the above expressions for which all
corrections to the EW observables coming from the $U(1)$
sector of the theory disappear in this limit. However some
corrections, such as $\rho_{*}$ vanish~(not including those coming
from the triplet VEV) in only the $R=1/2$ limit and are angle
independent. However, $R=1/2$ is inconsistent with the gauge
invariance of our Yukawa coupling (\ref{miniyuk}), and so these
corrections can never disappear simply from judicious choices of
the $U(1)$ charges.  However, since the triplet VEV corrections go
in the opposite direction of the $U(1)$ corrections it is not
necessary to have $R=1/2$ exactly.

\subsection{Numerical bounds}

\begin{figure}[t]
\centerline{\includegraphics[width=0.5\hsize]{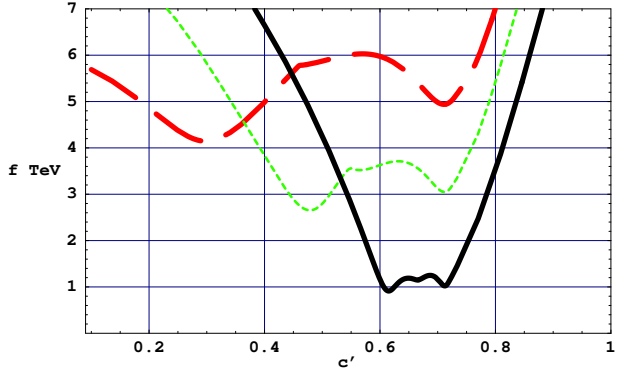}}
\caption{The 95\% confidence level bound is shown for $a=1$
and $R=1$ (dashed), $R=4/5$ (dotted), and $R=3/5$ (solid).
The bounds for
$R=2/5,1/5$ and $0$ can be obtained by reflecting the above curves
around $c'^2=1/2$ due to the $R\to 1-R$, $c'^2\to 1-c'^2$ symmetry
of the expressions for the EW corrections.}
\label{rfig}
\end{figure}
To analyze the bounds on the model we have performed a
three-parameter fit ($f,c,c'$) for the allowed values of
$R=1,4/5,3/5,2/5,1/5$ and $0$. This determines the gauge sector of
the model, but there is also the additional parameter $a$ in the
Coleman-Weinberg potential of the Littlest Higgs that affects the
size of the triplet VEV\@. The parameter $a$ is expected to be
${\cal O}(1)$; we consider fixed values of $a$ in the range
$0.5$-$5$. To ensure the high energy gauge couplings
$g_{1,2},g'_{1,2}$ are not too strongly coupled, the angles
$c,s,c^\prime,s^\prime$ cannot be too small.  As before
\cite{bigcorrections} we allow for $c,s,c',s' > 0.1$, or
equivalently $0.1 < c,c' < 0.995$. We allow $f$ to take on any
value (although for small enough $f$ there will be constraints
from direct production of $B_H$).  The general procedure we used
is to systematically step through values of $c$ and $c'$, finding
the lowest value of $f$ that leads to a shift in the $\chi^2$
corresponding to the 95\% confidence level (C.L.). For a
three-parameter fit, this corresponds to a $\Delta \chi^2$ of
about $7.8$ from the minimum. In Fig.~\ref{rfig} the allowed
values of $f$ are plotted as a function of $c$ for fixed values of
$R$ and $a$ and one sees that the value of allowed $f$ dips down
to $1\,\mbox{TeV}$ for $R=3/5$.
\begin{figure}[h!b]
\centerline{\includegraphics[width=.9\hsize]{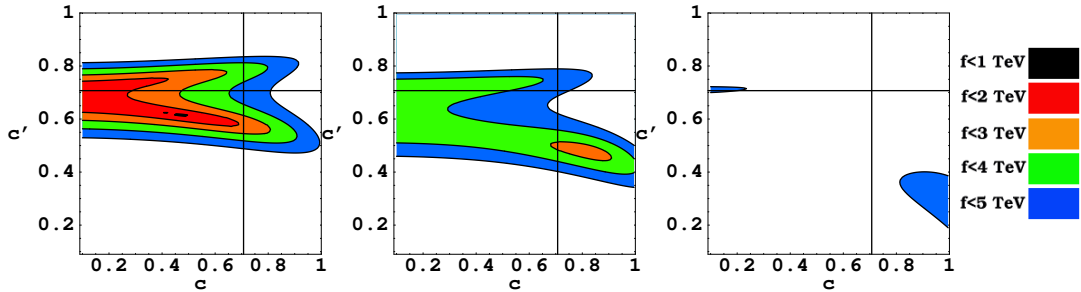}}
\caption{Contours of the minimum allowed values of $f$ at 95\% C.L.
for $a=1$ and $R=3/5$ (left graph) $R=4/5$ (center graph) $R=1$ (right graph).}
\label{fifths}
\end{figure}

In Fig.~\ref{rfig} the value of $c$ has already been chosen to
minimize the bound on $f$ so this only shows the size of allowed
parameter space in $c'$.  In Fig.~\ref{fifths} we plot for fixed
$R=3/5,4/5,1$ and $a=1$ a contour plot showing the allowed range
of parameter space at 95\% C.L. for both $c$ and $c'$ showing the
size of the allowed region of parameter space for a given value of
$R$ and $a$.  We see in Fig.~\ref{fifths} that the allowed region
where the bound on $f$ reaches $1$ TeV is extremely small.  So,
the Littlest Higgs with $R=3/5$ does have regions of parameter
space where the bound on $f$ is around $2$ TeV and thus the
fine-tuning in the Higgs mass is minimized.  Nevertheless, the
size of this region is not particularly large for $f \le 2$ TeV, and
for $f \le 1$ TeV it is essentially just a point in parameter space.

\begin{figure}[h!b]
\centerline{\includegraphics[width=0.7\hsize]{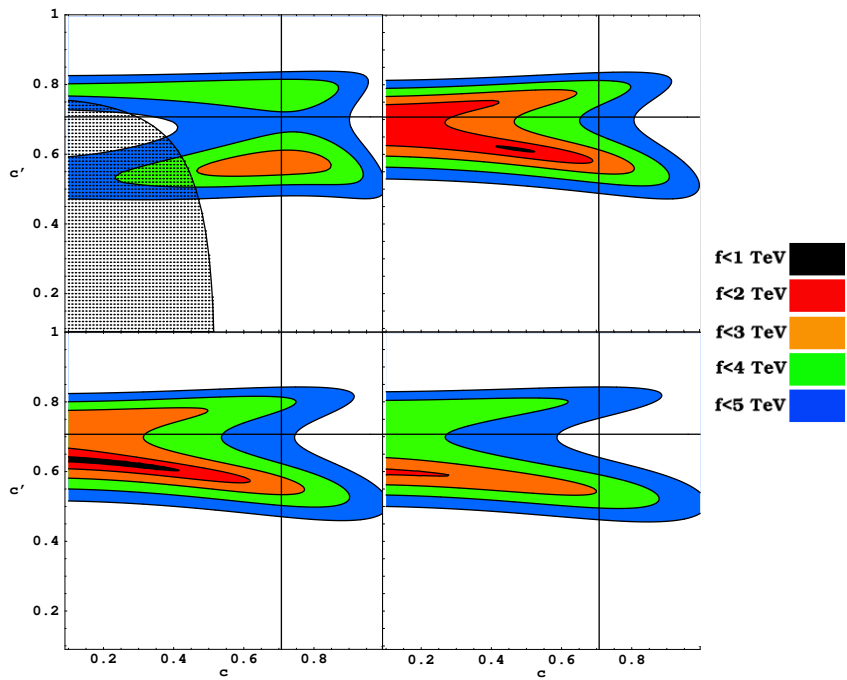}}
\caption{Plot of allowed values of $f$ at $95$ \% C.L. for
$a=0.5,1,2,5$ starting from the upper left corner in respective
clockwise order, and fixed $R=3/5$.  The shaded grey region
corresponds to the parameter space excluded due to a negative
triplet (mass)$^2$.  We do not show $a=0.1$ since virtually the entire
region is excluded due to a negative triplet (mass)$^2$.}
\label{aplot}
\end{figure}

Next, we consider how varying the value of $a$ affects the minimum
value of $f$ and the size of the allowed parameter regions at 95\%
C.L. Since the parameter $a$ not only affects the size of the VEV
but also feeds into the triplet mass, there is an additional
constraint on $a$ upon requiring a positive triplet (mass)$^2$.
If the triplet (mass)$^2$ were negative the triplet would obtain a
VEV of order $f$ that would introduce zeroth order corrections to
precision EW observables, and is thus ruled out.
Following the same procedure as outlined above, we recalculate the
bounds varying $a$ discretely between $a=0.5$ to $a=5$ and show
the allowed regions at 95\% C.L. in Fig.~\ref{aplot}. The
additional shaded areas are those excluded by ensuring the triplet
(mass)$^2$ is positive.

Fig.~\ref{aplot} shows that perturbing $a$ from one generally
reduces the allowed region of parameter space for a given
$f$. For small values of $a$ the exclusion region due to requiring
a positive triplet (mass)$^2$ is large.
\begin{figure}[h!t]
\centerline{\includegraphics[width=0.8\hsize]{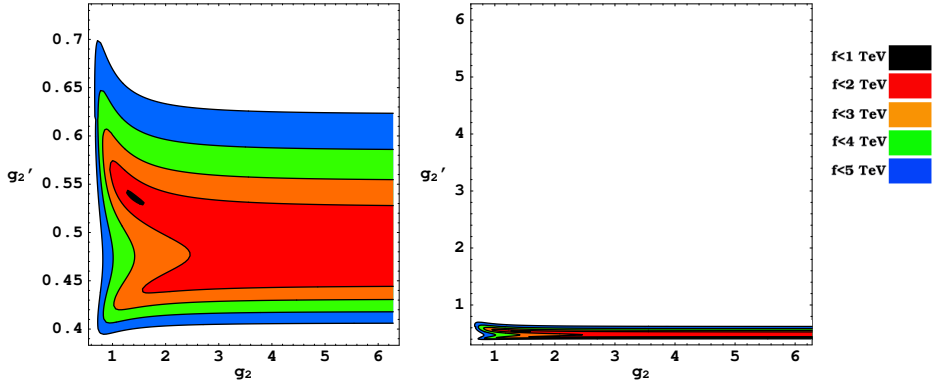}}
\caption{
The plot on the left corresponds to the phenomenologically viable
region of parameter space in $g_2$ and $g_2^\prime$ between $0$
and $2\pi$ with allowed values of $f$ at 95\% C.L.. The plot on
the right corresponds to showing all possible values of $g_2$ and
$g_2^\prime$ between $0$ and $2\pi$ with allowed values of $f$ at
95\% C.L.} \label{couplingplot}
\end{figure}
In Fig.~\ref{couplingplot} we demonstrate the allowed region of
parameter space in the actual couplings $g_2$ and $g_2^{\prime}$.
Due to the non-linear mapping between the angles and the physical
couplings we see that the allowed region in the physical coupling
space is quite small.  Similarly, plotting in the space of $g_1$
and $g_1^{\prime}$ one finds the allowed regions are even smaller.
Plotting the physical couplings illustrates that the small but
finite extent of the allowed region in $c$ and $c'$ space is
even further suppressed in the physical coupling space.
\par
To demonstrate the significance of the bounds on $f$ we quantify
the amount of fine tuning necessary as done in~\cite{littlest}.
The heavy fermion (with mass $m^\prime$) introduced to cancel the
top loop divergence of the SM contributes
\begin{equation}\label{ft}
- \frac{3 \lambda_t^2}{4 \pi^2} m^{\prime 2} \log \frac{4 \pi
f}{m^\prime}
\end{equation}
to the Higgs mass squared.  Given the bound on the mass of the
heavy fermion $m^\prime > \sqrt{2} f$ as calculated
in~\cite{bigcorrections} we can calculate the percentage of
fine-tuning required for a given value of $f$.  In
Fig.~\ref{fifths} the contours are plotted for the regions
$f<(1,2,3,4,5)$ TeV and $f>5$ TeV, this corresponds to a fine
tuning (assuming a physical Higgs mass of order 200 GeV)\footnote{The
  plots above assume $\lambda=1/3$, corresponding to a 200 GeV Higgs,
  while the fits perturb the SM with a $115$~GeV Higgs.  However, because the triplet VEV
  depends only on the ratio $\lambda/a$, with $a$ appearing nowhere
  else, one can simply scale $a$ in
  the plots above by a
  factor of $\sim 1/3$ to get the $115$~GeV results.}   
  of more than $(14.3,3.6,1.6,0.9,0.6)\%$ and less than
$0.6\%$ respectively. One can see from Fig.~\ref{fifths} that
there \emph{are} regions in parameter space where the fine-tuning
is on the order of 10\%
but they are very small.
%%%%%%%%%%%%%%%%%%%%%%%%%%%%%%%%%%%%%%%%%%%%%%%%%%%%%%%%%%%%%%%%%%%%%%%%%%%%%%%%%%%%%%%%%%%%%%%%%%%%
%%%%%%%%%%%%%%%%%%%%%%%%%%%%%%%%%%%%%%%%%%%%%%%%%%%%%%%%%%%%%%%%%%%%%%%%%%%%%%%%%%%%%%%%%%%%%%%%%%%%%
\section{Modifying the Littlest Higgs}
\label{modlittlest}\setcounter{equation}{0}

In the Littlest Higgs model, the largest contributions to the
EW precision observables come from the $U(1)$ sector.
We have already seen that modifying the light fermion charges
is one alternative that can relax the bounds on the model.
Here we examine two possibilities for modifying the $U(1)$ sector:
choosing $U(1)$'s that are not necessarily subgroups of the global $SU(5)$;
and, gauging only $U(1)_Y$~\cite{Nima}.  We will examine the benefits
and drawbacks of each of these two possibilities.

\subsection{Changing the gauged U(1)'s}\label{bmodsec}

If we do not insist that the gauged $U(1)$'s be $SU(5)$ subgroups
then we can introduce a new parameter $b$ with which the $U(1)$'s
are modified to \begin{equation} Y_1=\mbox{diag}\,(-3,-3,2,2,2)/10
+ bI, \end{equation} \begin{equation}
Y_2=\mbox{diag}\,(-2,-2,-2,3,3)/10 - bI, \end{equation} where $I$
is the $5$ by $5$ unit matrix.  This modification of the $U(1)$
sector does not change the $U(1)_Y$ quantum numbers and can be
effectively thought of as a way to decouple the heavy $U(1)$ gauge
bosons. This introduces an extra singlet into the Goldstone boson
matrix, whose effects we ignore.  One can easily see that that for the cases $b=1/5$ or
$b=-1/5$ that the $U(1)$'s are orthogonal. This would be a
preferred value of $b$, since then the issue of loop induced
kinetic mixing terms would not arise.  When computing the bare
expressions in this modified  model we find for instance that
\begin{equation} \label{newrhost}
\rho_{*}=1+\frac{5}{4}(1-2R)^2\Delta\left(\frac{1}{1+100b^2}\right)-4\Delta'
\end{equation}
compared with
\begin{equation}
\rho_{*}=1+\frac{5}{4}(1-2R)^2\Delta-4\Delta'
\end{equation}
in the Littlest Higgs model.  Here $R$ is again characterizing the
embedding of the fermions into $U(1)_1\times U(1)_2$ as in the
previous section. The $1/(1+100b^2)$ factor suppresses all other
$U(1)$ dependent expressions as well and thus naively would be
thought of as a very efficient way to get rid of the bound on $f$.
The reason is that for large values of $b$ the heavy $U(1)$ gauge
boson would live mostly in the $U(1)$ which is not part of $SU(5)$
(the extra $U(1)$ of $U(5)$) which does not violate custodial $SU(2)$.
The effect of the new $U(1)$ on other quantities is similar and
is as follows:
\begin{eqnarray}
M_W^2&=& \frac{g^2
v^2}{4}\left(1+\frac{\Delta}{4}(1+4c^2-4c^4)+4\Delta'\right)\\
M_Z^2&=&
\frac{(g^2+g'^2)v^2}{4}\left(1+\frac{\Delta}{4}(1+4c^2-4c^4)-\frac{\Delta 5(1-2c'^2)^2}{4(1+100b^2)}+8\Delta'\right)\\
s_{*}^2(0)&=&s_W^2+\frac{\Delta}{2}\left(-c^2
s_W^2+\frac{5c_W^2}{1+100b^2}(-1+c'^2+R)(2R-1)\right)\\
s_0^2&=&s_W^2-\frac{s_W^2c_W^2}{c_W^2-s_W^2}\left[\Delta\left(c^2-c^4-\frac{5}{4(1+100b^2)}(1-2c'^2)^2\right)+4\Delta'\right]\\
G_F&=&\frac{1}{\sqrt{2}v^2}\left(1-\frac{\Delta}{4}-4\Delta'\right)
\end{eqnarray}

However, we need to make sure that the Yukawa couplings remain
invariant under the modified $U(1)$ charge assignments. In order for
the operator giving rise to the top Yukawa coupling in
(\ref{topyuk}) to be invariant we need the relation
\begin{equation}
R=\frac{3+40b}{5}
\end{equation}
to be satisfied. If we insist that all three families have the
same charge assignments (that is that all the one loop quadratic
divergences could be canceled at least in principle) then this
relation needs to hold for the first two families as well. In
order to get
 $b=\pm 1/5$ for orthogonal
$U(1)$'s we must have $R=11/5$ and $R=-1$ respectively.  For such
large values of $R$ the corrections to EW observables
would in fact increase, not decrease. For example one can see
that the expression for $\rho_*$ does not go to 1 if we increase
$b$, but rather asymptotes to $1+ 16/5 \Delta$ for large $b$. Thus
very large values of $b$ (contrary to our original expectation)
will not be helpful. However, one can still use this freedom to
slightly relax the bounds obtained in the previous section. The
reason is that by picking $b=-\frac{1}{80}$ one can set
$R=\frac{1}{2}$, in which case the heavy $U(1)$ gauge boson
contributions to EW precision observables can almost all be
eliminated, its coupling to the fermions can be eliminated. Previously this
point was not allowed as long as we restricted ourselves to
operators with integer powers of fields.
\begin{figure}[h!t]
\centerline{\includegraphics[width=.5\hsize]{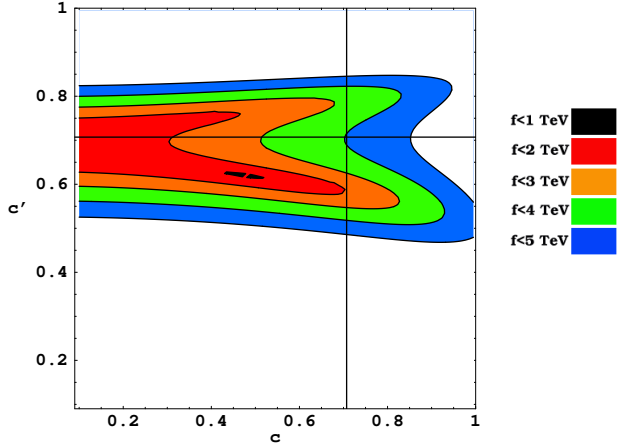}}
\caption{Contour plot of allowed values of $f$ as a function of
the angles $c$ and $c'$ for $R=1/2$ and $a=1$ and $b=-1/80$.}
\label{fig:u1bmod}
\end{figure}
The region allowed at the 95\% CL by the EW precision fit
for the $R=1/2, b=-1/80$ case is shown in Fig.~\ref{fig:u1bmod}.
We can see that the region corresponding to the smallest fine
tuning ($f<2$ TeV) is slightly, but not significantly,  larger
than in the cases considered in the previous section, and there is
now a non-negligible region where $f$ would be less than $1$ TeV\@.

\subsection{Gauging only U(1)$_Y$}
Since many of the corrections actually originate from the exchange
of the heavy $U(1)$ gauge boson, one way to try to improve the
EW precision bounds is by eliminating the heavy
$U(1)$ and only gauging the SM $U(1)_Y$ subgroup~\cite{Nima}.  This
again leaves an extra scalar in the $\Pi$ matrix whose effects we
ignore.  In the case where only one $U(1)_Y$ is gauged all constant
corrections in $\Delta$ and $U(1)$ angle dependent pieces in our
bare expressions disappear. We calculate all the corrections from
the kinetic term which is \beq \frac{f^2}{8} {\rm Tr} D_\mu \Sigma
(D^\mu \Sigma)^\dagger \eeq where now \beq D_\mu \Sigma =
\partial_\mu \Sigma - i \sum_j \left[ g_j W_j^a (Q_j^a \Sigma +
\Sigma Q_j^{aT} )\right]- i g^\prime B( Y_{SM} \Sigma + \Sigma
Y_{SM})~, \eeq where $g_j$ is the coupling of the $[SU(2)]_j$
gauge group, and $g^\prime$ is the coupling of the $U(1)_Y$ gauge
group.  The generator for $U(1)_Y$ is
$Y_{SM}=\mbox{diag}(-1/2,-1/2,0,1/2,1/2)$. Recalculating the
relevant quantities for this case yields:
\begin{eqnarray}
M_W^2&=& \frac{g^2
v^2}{4}\left(1+\frac{\Delta}{4}(1+4c^2-4c^4)+4\Delta'\right)\\
M_Z^2&=&
\frac{(g^2+g'^2)v^2}{4}\left(1+\frac{\Delta}{4}(1+4c^2-4c^4)+8\Delta'\right)\\
s_{*}^2(0)&=&s_W^2+\frac{\Delta}{2}\left(-c^2
s_W^2\right)\\
s_0^2&=&s_W^2-\frac{s_W^2c_W^2}{c_W^2-s_W^2}\left[\Delta\left(c^2-c^4\right)+4\Delta'\right]\\
G_F&=&\frac{1}{\sqrt{2}v^2}\left(1-\frac{\Delta}{4}-4\Delta'\right)\\
\rho_{*}&=&1-4\Delta'
\end{eqnarray}
We see from these expressions we still retain our constant shifts
in $\Delta'$ to our bare observables.  We must compute the scalar
Higgs potential coming from the quadratically divergent pieces of
the Coleman-Weinberg(CW) potential so as to estimate the size of
the triplet VEV in this particular model. The quadratically
divergent piece of the CW potential from the gauge bosons is
\begin{equation}
\frac{\Lambda^2}{16\pi^2}\mbox{Tr}M_V^2(\Sigma),
\end{equation}
where $M_V^2$ is the gauge boson mass matrix in an arbitrary
$\Sigma$ background.  $M_V^2$ can be read off from the covariant
derivative for $\Sigma$, giving a potential
\begin{equation}
ag_j^2
f^4\sum_a\mbox{tr}\left[(Q_j^a\Sigma)(Q_j^a\Sigma)^*\right]+ag^\prime
f^4\mbox{tr}\left[(Y_{SM}\Sigma)(Y_{SM}\Sigma)^*\right]
\end{equation}
where $a$ is a constant of order one determined by the relative
size of the tree-level and loop induced terms, and we have used
$\Lambda\sim 4\pi f$.  We also have a quadratically divergent
contribution to the CW potential coming from the fermion loops.
This potential from the fermion sector contributes the same
potential as that generated from the $SU(2)_2$ gauge bosons since
the operator that gives the fermion potential is $SU(3)_1$
symmetric.  Calculating the VEV of $\phi$ we obtain
\begin{equation}
v^\prime=\frac{iv^2}{4f^2}\frac{a^\prime
\lambda_1^2+a(g_1^2-g_2^2)}{-a^\prime\lambda_1^2+a(4g^{\prime
2}+g_1^2+g_2^2)}
\end{equation}
where $a^\prime$ is an order one coefficient parameterizing the
fermion operator that contributes to the CW potential and
$\lambda_1$ is the Yukawa coupling of that operator.  We can
integrate out the triplet to generate a quartic coupling for the
Higgs
\begin{equation}\label{quarticc}
\lambda=\frac{a(-a^\prime \lambda_1^2(2g^{\prime 2}+3 g_1^2)+a(-4
g^{\prime 4}+3 g_1^2 g_2^2+2 g^{\prime
2}(g_1^2+g_2^2)))}{3(-a^\prime \lambda_1^2+a(4 g^{\prime
2}+g_1^2+g_2^2))}.
\end{equation}
and thus we obtain that
\begin{equation}\label{deltap}
\Delta^\prime \equiv \frac{|v^\prime|^2}{v^2}=
\frac{\Delta}{144}\left[1+\frac{6\lambda-4a g_1^2}{a(2g^{\prime
2}+g_1^2)}\right]^2.
\end{equation}
There is one further constraint that requires the triplet mass to
be positive so the triplet does not obtain a VEV at the scale $f$.
The positivity of the triplet mass requires
\begin{equation}
a(4g^{\prime 2}+g_1^2+g_2^2)-a^\prime \lambda_1^2 >0
\end{equation}
which is equivalent to [using (\ref{quarticc})]
\begin{equation}
2a g^{\prime 2} -3\lambda + 3 a g_1^2>0.
\end{equation}
This constraint combined with (\ref{deltap}) gives us that
\begin{equation}
\Delta^\prime < \frac{\Delta}{16}.
\end{equation}
It is important to note that even though $\Delta^\prime$ is small,
the coefficients of $\Delta^\prime$  in the contributions to observables are constant
and can not be fully tuned away except in small regions of
parameter space.
\par
Using the bare expressions and re-expressing them in terms of
observables we are able to calculate the bounds on the Littlest
Higgs when we only gauge $U(1)_Y$.
\begin{figure}[h!tb]
\centerline{\includegraphics[width=0.5\hsize]{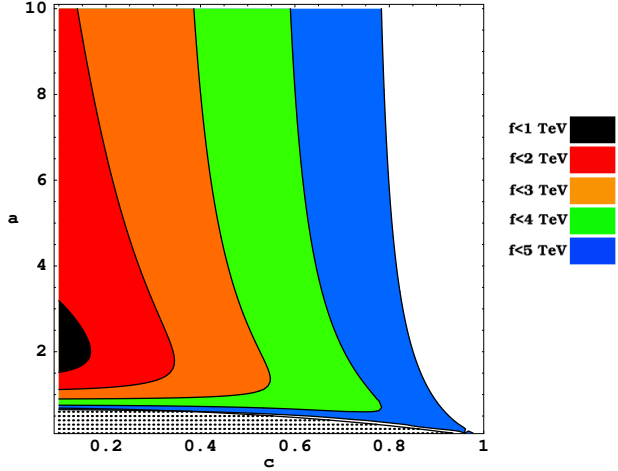}}
\caption{Contour plot of the allowed values of $f$ as a function
of $a$ and $c$.  The x-axis is $c$ and the y-axis is $a$.  The
grey shaded region at the bottom corresponds to the region
excluded by requiring a positive triplet mass.} \label{oneu1a}
\end{figure}
In Fig.~\ref{oneu1a} we have plotted the allowed values of $f$ as
a function of $c$  which parameterizes the $SU(2)$ mixing and $a$
which parameterizes the strength of the gauge contribution to the
CW potential.  From Fig.~\ref{oneu1a} we see that in a relatively
small region of parameter space the bounds on $f$ make the model
acceptable from a fine tuning perspective using the measure of
fine tuning defined as (\ref{ft}).

\begin{figure}[h!tb]
\centerline{\includegraphics[width=0.5\hsize]{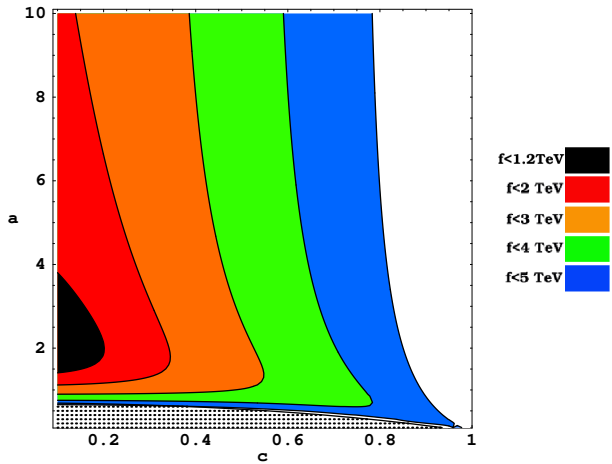}}
\caption{Shown above is a contour plot with only one $U(1)$ gauged
in
  a variation on the Littlest Higgs model.  We have chosen a
  different lowest contour because the region with the lowest $f$
  is shifted slightly above $1$~TeV\@.
} \label{BSR}
\end{figure}

In the region of small $f$, the triplet vev contributions are
negligible, and the $SU(2)$ corrections are small due to
suppression by factors of $c$, thus one should ask how the
inclusion of loop contributions from the additional particles to $T$ and $\rho_*(0)$ might
affect the location and size of this region.  The largest contribution
is coming from the additional heavy vector-like top quarks. The leading
contribution
beyond what is induced by the normal SM loops is then given
by
\begin{equation}
\Delta \rho_{\mathrm{top}} = \frac{ N_c \Delta}{16 \pi^2} \left(
\frac{ \lambda_t^4 }{2 \lambda_H^2} \right) \log \left[ \frac{ 2
  \lambda_H^2 }{\lambda_t^2 \Delta} \right],
\label{rhotop}
\end{equation}
where $\lambda_H\equiv\sqrt{\lambda_1^2+\lambda_2^2}$, with
$\lambda_1$ and $\lambda_2$ defined in (\ref{topyuk}).  Note that $\lambda_H$ has a
minimum value of $\lambda_H=\sqrt{2}$. One can see that this
contribution decouples for $f\to \infty$, and is not important for
establishing the bounds as long as the tree-level contributions are
already forcing $f$ to be large. However, for the regions with small
$f$ these contributions are comparable to the non-canceled pieces of
the tree-level effect, and thus might be relevant when one is trying
to find the precise shape of the allowed regions.
(\ref{rhotop}) can be interpreted as
a $T$ contribution for the $Z$-pole observables, and as a
correction to $\rho_* (0)$ in the low energy observables. The
results of including the maximal shift are shown in
Fig.~\ref{BSR}. Note that the regions where $f$ goes below $1$~TeV
are slightly shifted above $1$~TeV\@.

The drawback of this model is that without gauging a second $U(1)$
the one-loop quadratic divergences to the Higgs mass from $U(1)_Y$
do not cancel. Therefore in the case of gauging only $U(1)$ we
will introduce divergences of the form
\begin{equation}
\frac{1}{16\pi^2}g'^2\Lambda^2.
\end{equation}
For $\Lambda\sim 10\,\mbox{TeV}$ this implies a fine tuning of the
Higgs mass of approximately $50$ percent, which is obviously
preferable to the level of fine tuning that the original
little Higgs models generically had.  However, in this case we are
giving up on the concept of systematically eliminating all
one-loop quadratic divergences arising from interactions with
order one coefficients.

%%%%%%%%%%%%%%%%%%%%%%%%%%%%%%%%%%%%%%%%%%%%%%%%%%%%%%%%%%%%%%%%%%%%%%%%%%%%%%%%%%%%%%%%%%%%%%%%%%%%
%%%%%%%%%%%%%%%%%%%%%%%%%%%%%%%%%%%%%%%%%%%%%%%%%%%%%%%%%%%%%%%%%%%%%%%%%%%%%%%%%%%%%%%%%%%%%%%%%%%%%

\section{SU(6)/Sp(6) Little Higgs}
\label{su6sp6sec}\setcounter{equation}{0}

To determine whether the bounds on $f$ are generically large in
little Higgs coset models, here we consider the $SU(6)/Sp(6)$ model
proposed in~\cite{su6sp6}.

\subsection{The SU(6)/Sp(6) model}

This model is based on an anti-symmetric
condensate breaking a global $SU(6)$ to $Sp(6)$ in contrast to the
symmetric condensate breaking used in~\cite{littlest}.
The basis for the breaking is characterized by the direction $\Sigma_0$
which is given by
\begin{equation}
\langle \Sigma_0 \rangle = \left(
\begin{array}{cc}
0 & -I\\
I & 0\\
\end{array}\right),
\end{equation}
where $I$ is the $3$ by $3$ unit matrix. The Goldstone bosons are
then described by the pion fields $\Pi=\pi^a X^a$, where the $X^a$
are the broken generators of the $SU(6)$.  The non-linear sigma
model field is then
\begin{equation}
\Sigma=e^{i\Pi /f}\Sigma_0 e^{i\Pi^T/f}=e^{2 i \Pi/f}\Sigma_0,
\end{equation}
where $f$ is the scale of the VEV that accomplishes the breaking.
An $SU(2)^2$ portion of the $SU(6)$ global symmetry is gauged,
where the generators are
\begin{equation}Q_1^a = \frac{1}{2}
\left(
\begin{array}{c|c}
\begin{array}{c|c}
\mbox{\Large $\sigma^a$} & 0 \\ \hline
0 & 0 \\
\end{array} & \;\;\;\mbox{\Large $0$} \\ \hline
\, & \, \\
\mbox{\Large $0$} & \;\;\;\mbox{\Large $0$} \\
\, & \, \\
\end{array}
\right) \hspace{30pt} {\rm and} \hspace{30pt}
Q_2^a=-{\frac{1}{2}}\left(
\begin{array}{c|c}
\, & \, \\
\mbox{\Large $0$} & \;\;\;\mbox{\Large $0$} \\
\, & \, \\ \hline
 \;\;\;\mbox{\Large $0$} &\begin{array}{c|c}
\, & \, \\
\mbox{\Large $\sigma^{a*}$} & 0 \\ \hline
0 & 0 \\
\end{array}\\
\end{array}
\right),
\end{equation}
where $\sigma^a$ are the Pauli $\sigma$ matrices.  There are also
two gauged $U(1)$'s that are not a subgroup of the global $SU(6)$
which are given by
\begin{equation}
Y_1=\mbox{diag}(0,0,-1/2,0,0,0)
\end{equation}
and
\begin{equation}
Y_2=\mbox{diag}(0,0,0,0,0,1/2).
\end{equation}
The Goldstone boson matrix $\Pi$ is expressed in terms of the
uneaten fields as
\begin{equation}
\Pi=\left(
\begin{array}{c|c|c|c}
\mbox{\LARGE $0$} & \frac{\phi_1}{\sqrt{2}} &
\begin{array}{cc}
0 & s \\
-s & 0 \\
\end{array} & \frac{\phi_2}{\sqrt{2}}  \\ \hline
\phi_1^\dagger / \sqrt{2} & 0 & -\phi_2^T /\sqrt{2} & 0 \\ \hline
\begin{array}{cc}
0 & -s^* \\
s^* & 0 \\
\end{array} &
-\frac{\phi_2^*}{\sqrt{2}} & \mbox{\LARGE $0$} & \frac{\phi_1^*}{\sqrt{2}} \\
\hline
\phi_2^\dagger / \sqrt{2} & 0 & \phi_1^T /\sqrt{2} & 0 \\
\end{array}
\right).
\end{equation}
In this model $\phi_1$ and $\phi_2$ transform under $SU(2)_W$ and
$U(1)_Y$ as $\mathbf{2}_{1/2}$ and $\mathbf{2}_{-1/2}$
respectively, whereas $s$ transforms as $\mathbf{1}_0$.  The
kinetic term of the non-linear sigma model field is
\begin{equation}
-\frac{f^2}{8}\mbox{Tr}[(D_\mu \Sigma)(D^\mu \Sigma)^*]
\end{equation}
where
\begin{equation}
D_\mu\Sigma=\partial_\mu \Sigma -\sum_{j}[i g_j W_j^a(Q_j^a
\Sigma+\Sigma Q_j^{aT})+i g_j' B_j(Y_j \Sigma + \Sigma Y_j^T)]
\end{equation}
where $g_j$ and $g_j'$ are the couplings of $[SU(2)\times U(1)]_j$
gauge groups.  Following the same procedure as in~\cite{bigcorrections} we
expand the kinetic term in $\Sigma$ to second-order. (In this model
we have to expand to second order since we have two Higgs doublets
and can not redefine just one VEV when expanding from linear to
higher orders as could be done in the Littlest Higgs model.) The
heavy gauge bosons acquire masses \beq
M_{W_H} &=& \sqrt{g_1^2 + g_2^2} \frac {f}{2}~,\\
M_{B_H} &=& \sqrt{ g_1^{\prime 2} + g_2^{\prime 2}} \frac {f}{2
\sqrt{2}}~. \eeq
Note, that contrary to the Littlest Higgs model the heavy $U(1)$ gauge
boson is not significantly lighter than the $SU(2)$ gauge bosons,
which itself reduces the bounds slightly.
 The
VEV's the $\phi$'s acquire are
\begin{equation}
\langle\phi_1\rangle=\frac{1}{\sqrt{2}} \left(
\begin{array}{c}
0\\
v_1\\
\end{array}
\right) \mbox{ and } \langle\phi_2\rangle =
\frac{1}{\sqrt{2}}\left(
\begin{array}{c}
v_2\\
0\\
\end{array}\right),
\end{equation}
which we will redefine in terms of
\begin{equation}
v^2=v_1^2+v_2^2
\end{equation}
and
\begin{equation}
\tan \beta=\frac{v_2}{v_1}.
\end{equation}
The singlet which is very heavy will also get a VEV however it is
neutral and does not enter into any of our expressions.  We will
also make further simplifications by defining
\begin{equation}
\Delta=v^2/f^2,
\end{equation}
\begin{equation}
s=\frac{g_2}{\sqrt{g_1^2+g_2^2}},c=\frac{g_1}{\sqrt{g_1^2+g_2^2}},
s'=\frac{g_2'}{\sqrt{g_1^{\prime 2}+g_2^{\prime 2}}},
c'=\frac{g_1' }{\sqrt{g_1^{\prime 2}+g_2^{\prime 2}}},
\end{equation}
\begin{equation}
g=g_1 s, g'=g_1' s', \;\mbox{ and }
s_W=\frac{g^\prime}{\sqrt{g^{2}+g^{\prime 2} } }
\end{equation}

\subsection{Contributions to electroweak observables}

We assume all left-handed fermions are charged only under $SU(2)_1$.
Analogous to our discussion in a variation of the Littlest Higgs model,
we allow fermions to couple to both $U(1)$'s with a constant $R$ varying
from $0$ to $1$ characterizing the coupling to $U(1)_1$ ($R=1$) and
$U(1)_2$ ($R=0$).  Below we will determine the restrictions on the
parameter $R$ by requiring gauge invariance of the Yukawa couplings.

Expanding the relevant bare expressions to
first order in $\Delta$ we find
\begin{eqnarray}
M_W^2&=&g^2
\frac{v^2}{4}\left(1-\frac{\Delta}{4}(1-4c^2+4c^4-\cos^2
2\beta)\right) \nonumber \\
M_Z^2&=&(g^2+g^{\prime 2})\frac{v^2}{4}\left(1-\frac{\Delta}{4}(1-4c^2+4c^4+2(s^{\prime 2}-c^{\prime 2})^2)\right)\nonumber \\
G_F&=&\frac{1}{\sqrt{2}v^2}\left(1+\frac{\Delta}{4} (1-\cos^2
2\beta)\right)\nonumber \\
\delta \tilde g_{ff} &=& T^3 \frac{\Delta}{2} \left(-2 + c^2 -
2\,c^4 + 6\,c^{\prime 2} - 4\,c^{\prime 4} + 2\,R -
  4\,c^{\prime 2}\,R \right) \nonumber \\&& + Q \Delta \left( -1 + 2\,c^{\prime 2}  \right) \,
     \left( -1 + c^{\prime 2}  + R \right) \nonumber
\end{eqnarray}
\begin{eqnarray}
s_0^2 &=& s_W^2+ \frac{s_W^2 c_W^2}{c_W^2-s_W^2}  \frac{\Delta}{4}
\left[ 2 - 4\,c^2 + 4\,c^4 - 8\,c^{\prime 2} +   8\,c^{\prime 4}
   + \,\cos^2 2\beta  \right]\nonumber \\
 \rho_{*}&=&1+\frac{\Delta}{4}(2-8R+8R^2+\cos^2
2\beta)\nonumber \\
s_{*}^2(0)&=&s_W^2+\frac{\Delta}{2}\left(-s_W^2 c^2+2
c_W^2(c^{\prime 2}-1+R)(2R-1)\right)
\end{eqnarray}
It is clear that there is a new source of isospin breaking in this model when
$\cos^2 2\beta \ne 0$.  This is a generic result for non-linear $\sigma$ models with
two Higgs doublets \cite{Sekhar}.

\subsection{Numerical results}

The bare expressions given above can be used
to calculate all the EW observables in this model.
(For definitions see Ref. \cite{bigcorrections}.) We first
re-express the observables in terms of inputs $G_F$, $M_Z$, and
$\alpha(M_Z^2)$. The resulting expressions are then used to obtain
the bounds on $f$.  In the case of the ``minimal'' $SU(6)/Sp(6)$
model~\cite{su6sp6}, $R=1$ (all fermions coupling to $SU(2)_1 \times U(1)_1$).
Requiring gauge invariance of the Yukawa couplings (analogously to what
we did for the Littlest Higgs in Section~\ref{u1little}) we find that
$R=n$ where $n$ is any integer.  Again, we allow
$0.1 < c,c' < 0.995$, stepping through values of $c$, $c'$ and
$\tan \beta$, finding the
lowest value of $f$ that leads to a shift in the $\chi^2$
corresponding to the 95\% confidence level.  For a
four-parameter fit, this corresponds to a $\Delta \chi^2$ of about
$9.5$ from the minimum.
\begin{figure}[h!t]
\centerline{\includegraphics[width=0.7\hsize]{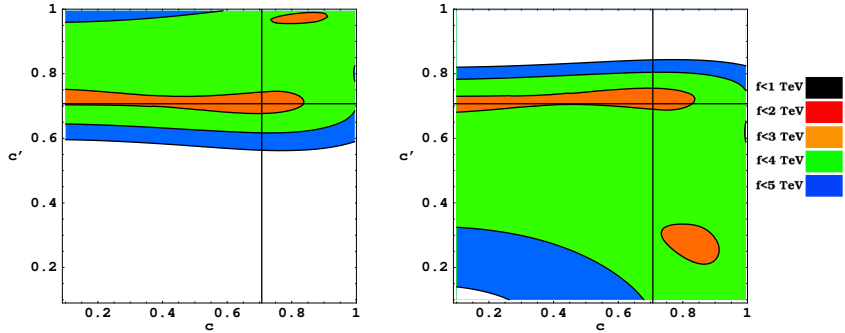}}
\caption{Contour plot of allowed values of $f$ as a function of
the angles $c$ and $c'$ for $R=0$ and $R=1$ with the 95\% C.L.
shown.} \label{su6sp6plt}
\end{figure}
In Fig.~\ref{su6sp6plt} we plot the allowed contour levels at the
$95\%$ C.L. for $R=0$, the point where in principle all quadratic
divergences could be canceled, and $R=1$ the smallest number of
insertions of extra fields required by our ``R" rule.  We find
that for neither $R=0$ or $R=1$ the bounds on this model show
significant improvement from the Littlest Higgs model.  In the
majority of parameter space as can be seen from
Fig.~\ref{su6sp6plt} the bounds on $f$ are greater than $3$ TeV\@.
This bound is surprising at first since the triplet was shown to
have a large influence when we examined the bounds on the Littlest
Higgs with gauging only $U(1)_Y$.  Without the triplet we have no
way of canceling off the large $U(1)$ corrections that contribute
a shift in $\Delta$ that has the opposite sign of the triplet
corrections in the Littlest Higgs.  Therefore in the most minimal
version of the $SU(6)/Sp(6)$ model it has the same problem as the
Littlest Higgs i.e. the large contributions from the $U(1)$ sector
of the theory.  We do not compute the fine tuning in the
$SU(6)/Sp(6)$ model since the contributions to the the Higgs mass
depend on a large number of Yukawa couplings that are not
constrained as in the Littlest Higgs.  In principle the large
bound on $f$ does not directly imply a large fine tuning since a
choice of parameters could reduce the level of fine tuning however
this specific choice of parameter would have to be explained in a
UV completion.

\subsection{Modifying SU(6)/Sp(6)}

Given the relatively strong bounds on $f$ in the ``minimal'' cases
$R=1,0$, we now consider modifying the $U(1)$ sector of the model.
In Section~\ref{modlittlest} we demonstrated the effects of adding
an additional $U(1)$ and only gauging $U(1)_Y$.  With the extra
$U(1)$ contribution modification
the Littlest Higgs model was slightly improved since the $R=n/5$
rule could be violated, in particular to $R=1/2$ where the light
fermions decouple from the heavy $U(1)$ gauge boson.
Such special values of $R$ lowered the bounds on $f$ in this variation
of the Littlest Higgs model, but only in a rather
small region of parameter space.
The modification where only $U(1)_Y$ was gauged was also
not a very significant improvement since the triplet VEV could not be canceled
by a contribution from the $U(1)$ sector.  The allowed region of
parameter space where this modification of the Littlest Higgs model
had a low $f$ was again small since it corresponded directly to where
the triplet VEV vanished.  For $SU(6)/Sp(6)$, if only
hypercharge were gauged, there would be minimal corrections to
this model in a large region of parameter space since there is no
triplet VEV\@.  As before, this leaves open the question of why some
of the
Higgs quadratic divergences proportional to order one couplings
are not canceled (even though they may be numerically not very significant).
\begin{figure}[h!t]
\centerline{\includegraphics[width=0.5\hsize]{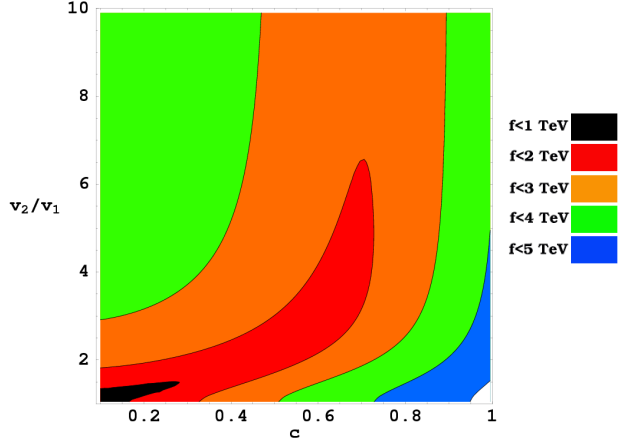}}
\caption{Contour plot of allowed values of $f$ as a function of
$c$ and $\tan\beta$ for gauging only $U(1)_Y$ in the $SU(6)/Sp(6)$
model with the $95\%$ C.L. shown} \label{su6oneu1plt}
\end{figure}
In Fig.~\ref{su6oneu1plt} we plot the allowed regions for the
$SU(6)/Sp(6)$ model when we gauge only $U(1)_Y$.  In this case we
see a dependence on the value of $\tan\beta$ since the large
$U(1)$ corrections do not dominate the fit that we perform as in
the minimal $SU(6)/Sp(6)$ case.  We see in Fig.~\ref{su6oneu1plt}
that gauging only $U(1)_Y$ enlarges the allowed region of
parameter space.
\par
In $SU(6)/Sp(6)$ the gauged $U(1)$'s are not subgroups of the
global $SU(6)$.  Since there is already a portion outside of the
the $SU(6)$ we could see the effect of slightly modifying the
$U(1)$'s in such a way as to preserve the approximate symmetries
they posses while minimizing the bounds on $f$.  The way we
accomplish this is through doing a $b$ modification as was
demonstrated for the Littlest Higgs in Section~\ref{bmodsec}.  In
the case of the $SU(6)/Sp(6)$ model the modification to the bare
parameters of the theory involving the $U(1)$ sector scale with a
factor $1/(1-8b+48b^2)$.  This scaling can be observed for
instance in the $\rho_{*}$ parameter where before
\begin{equation}
\rho_{*}=1+\frac{\Delta}{4}(2-8R+8R^2+\cos^2 2\beta)
\end{equation}
and with the $b$ modification
\begin{equation}
\rho_{*}=1+\frac{\Delta}{4}\left(\frac{2-8R+8R^2}{1-8b+48b^2}+\cos^2
2\beta\right).
\end{equation}
The modification involving $b$ is important because it allows
non-integer values of the $R$ parameter.  For the value $b=1/8$,
we find $R=1/2$, i.e. the fermions might be decoupled from the heavy $U(1)$
gauge boson.
\begin{figure}[h!t]
\centerline{\includegraphics[width=0.5\hsize]{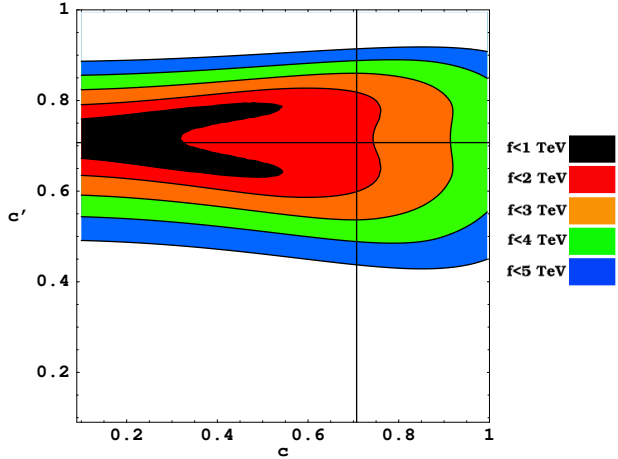}}
\caption{Contour plot of allowed values of $f$ as a function of
the angles $c$ and $c'$ for $R=1/2$ and $b=1/8$ with the $95\%$
C.L. shown.} \label{su6sp6pltrp5}
\end{figure}
In Fig.~\ref{su6sp6pltrp5} we demonstrate how using $b=1/8$ to
shift $R$ to $1/2$ changes the bounds on $f$ at the $95\%$ C.L..
As one can see from Fig.~\ref{su6sp6pltrp5} the bounds on $f$
decrease significantly, analogously to shifting from $R=1$ to
$R=1/2$ in the Littlest Higgs model.  Comparing the $b$
modification shown in Fig.~\ref{su6sp6pltrp5} for the
$SU(6)/Sp(6)$ model, and Fig.~\ref{fig:u1bmod} for the Littlest
Higgs model, we see that the $b$ modification opens up a larger
region of parameter space for $SU(6)/Sp(6)$ compared to the
Littlest Higgs.  The difference in allowed parameter regions in
the $b$ modification between the Littlest Higgs and the
$SU(6)/Sp(6)$ model can be attributed to the non-existence of a
triplet in the $SU(6)/Sp(6)$ model.  Since there are regions in
this model with $f<1$~TeV, to find the precise boundaries one
needs to calculate the loop corrections to observables in these
regions coming from the heavy top and the extra Higgs fields which is
done in~\cite{futuregregoire}.

%%%%%%%%%%%%%%%%%%%%%%%%%%%%%%%%
\section{SU(4)$^4$/SU(3)$^4$ Little Higgs}\label{dudes-sec}
\setcounter{equation}{0}
Finally, we consider a recently proposed little Higgs model based
on an $SU(4)\times U(1)$ gauge group~\cite{dudes}, where
$SU(2)_L$ is embedded into a simple group instead of $[SU(2)]^2$.
In this model there is a non-linear
sigma model for $[SU(4)/SU(3)]^4$ breaking with the diagonal $SU(4)$ subgroup
gauged and four non-linear $\sigma$ model fields $\Psi_i$ and $\Phi_i$
where $i=1,2$.  In order to be able to embed quarks into the
theory and to reproduce the SM value of $\sin^2\theta_W$, an
additional $U(1)$ group is needed. This
model appears to be a fundamentally different type of Little Higgs
model when compared to~\cite{littlest,su6sp6} due to the multiple
breaking of the global symmetry group by separate ``$\Sigma$''
fields.  This model is a little Higgs since it uses collective
symmetries to keep the Higgs a pseudo-Goldstone boson.  Instead of using
a product group, a simpler group is chosen and multiply broken by
four $\Sigma$ fields instead of one large breaking with a
single $\Sigma$. In addition, potential terms are added by hand rather
than being generated by gauge and Yukawa interactions.
This model could be viewed as the reincarnation
of the Higgs as a pseudo-Goldstone boson solution of the
doublet-triplet splitting problem of SUSY GUTs applied to Little
Higgs models~\cite{doublettriplet}. A convenient parameterization
of the non-linear $\sigma$ model fields in this model is
\begin{equation}
\Phi_1=e^{i \mathcal{H}_u \frac{f_2}{f_1}}\left(
\begin{array}{c}
0\\
0\\
f_1\\
0\\
\end{array}
\right)\;\; \Phi_2=e^{-i \mathcal{H}_u \frac{f_1}{f_2}}\left(
\begin{array}{c}
0\\
0\\
f_2\\
0\\
\end{array}
\right)
\end{equation}
\begin{equation}
\Psi_1=e^{i \mathcal{H}_d \frac{f_4}{f_3}}\left(
\begin{array}{c}
0\\
0\\
0\\
f_3\\
\end{array}
\right)\;\; \Psi_2=e^{-i \mathcal{H}_d \frac{f_3}{f_4}}\left(
\begin{array}{c}
0\\
0\\
0\\
f_4\\
\end{array}
\right)
\end{equation}
where
\begin{equation}
\mathcal{H}_u=\left(
\begin{array}{ccc}
\begin{array}{cc}
0 & 0 \\
0 & 0 \\
\end{array} &
h_u & \begin{array}{c} 0 \\ 0 \end{array}\\
h_u^\dagger & 0 & 0\\
\begin{array}{cc}
0 & 0
\end{array}
& 0 & 0
\end{array}
\right)/f_{12}\;\; \mathcal{H}_d=\left(
\begin{array}{ccc}
\begin{array}{cc}
0 & 0 \\
0 & 0 \\
\end{array} &
 \begin{array}{c} 0 \\ 0 \end{array}& h_d\\
\begin{array}{cc}
0 & 0
\end{array}
& 0 & 0\\
h_d^\dagger & 0 & 0\\
\end{array}
\right)/f_{34},
\end{equation}
and $f_{ij}=\sqrt{f_i^2+f_j^2}$.  The kinetic term for this model
is given by
\begin{equation}
\mathcal{L}_{kinetic}=|\mathcal{D}_\mu \Phi_i|^2+|\mathcal{D}_\mu
\Psi_i|^2+[\mbox{fermion and gauge kinetic terms}].
\end{equation}
where the gauge covariant derivative is
\begin{equation}
\mathcal{D}_\mu=(\partial_\mu + ig_4 A_\mu^a T^a - i \frac{g_X}{4}
A_\mu^X),
\end{equation}
and $A_\mu^a,g_4$ and $A_\mu^X,g_X$ are the gauge bosons and
couplings of the $SU(4)$ and $U(1)_X$ gauge groups respectively.
The diagonal generators for the $SU(4)$ group are
\beq
T^3&=&1/2\;\mbox{diag}(1,-1,0,0)~, \nonumber\\
T^{12}&=&1/2\;\mbox{diag}(0,0,1,-1)~,\nonumber\\
T^{15}&=&\frac{1}{\sqrt{8}}\;\mbox{diag}(-1,-1,1,1)~.
\eeq
Hypercharge
is a linear combination of the $U(1)_X$ generator and $T^{15}$,
$Y=\frac{1}{\sqrt{2}}T^{15}-I X$ where $I$ is the $4$ by $4$
identity matrix and $X$ is the $U(1)_X$ charge (for example, the $X$ charge
of the quark multiplets must be $5/12$ to give the correct $Y$ charge to
$u$ and $d$).  The charges of the two Higgs doublet fields
under $U(1)_Y$ are both $-1/2$ therefore the VEVs of the Higgs fields
are of the form
\begin{equation}
\langle h_u
\rangle=\frac{1}{\sqrt{2}}\left(\begin{array}{c}v_1\\0\end{array}\right)
\;\; \langle h_d
\rangle=\frac{1}{\sqrt{2}}\left(\begin{array}{c}v_2\\0\end{array}\right).
\end{equation}
We may simplify notation greatly by defining
\begin{equation}
v^2 = v_1^2+v_2^2, \;\;\:\: \tan \beta = \frac{v_2}{v_1},
\end{equation}
and
\begin{equation}
\tan \gamma = \frac{f_2}{f_1},\;\;\:\: \tan \eta =
\frac{f_4}{f_3},\;\;\:\: \tan \alpha = \frac{f_{34}}{f_{12}}.
\end{equation}
Here we choose to make the simplifying assumption
$f_{12}=f_{34}=f$.  This simplification retains the character of the
model with different $f_i$'s.
however we set the same scale for the VEV's of the $\Phi$'s and
$\Psi$'s.  The complete analysis with $f_{12}\neq f_{34}$ is
beyond the scope of this paper.
\par We expand the $\Phi_i$'s and $\Psi_i$'s to fourth order in
the Higgs fields and compute the bare expressions needed to
eventually calculate the shifts in the EW precision observables.
We make the identification that $g_4=g$ the coupling of the SM
$SU(2)$ group and
\beq
g'=g_X/\sqrt{1+\frac{g_X^2}{2g^2}}~.
\eeq

\subsection{Contributions to electroweak observables}

We find that the relevant bare expressions are
\begin{eqnarray}
\label{eqar:mart}
M_W^2 &=& \frac{g^2
v^2}{4}\left[1-\frac{\Delta}{6}\left(\cos^4\beta(\tan^2\gamma+\cot^2\gamma-1)
+\sin^4\beta(\tan^2\eta+\cot^2\eta-1)\right)\right] \nonumber \\
M_Z^2 &=& \frac{(g^2+g^{\prime 2})v^2}{4}\left[1-
\frac{\Delta}{6}\left(\cos^4\beta(\tan^2\gamma+\cot^2\gamma-1)
+\sin^4\beta(\tan^2\eta+\cot^2\eta-1)\right)
\right. \nonumber \\
&& \qquad\qquad\qquad\quad \left.
-\frac{\Delta}{4} \frac{(g^2-g'^2)^2}{g^4} \right] \nonumber \\
G_F&=&\frac{1}{\sqrt{2}v^2}\left[1+\frac{\Delta}{6} \left( \cos^4\beta(\tan^2\gamma+\cot^2\gamma-1)+\sin^4\beta(\tan^2\eta+\cot^2\eta-1) \right)
  \right. \nonumber \\
&&\left.-\frac{\Delta}{2}\left( \cot^2 \gamma  \cos^2 \beta + \tan^2 \eta \sin^2 \beta\right) \right] \nonumber \\
s_0^2&\equiv& \sin^2\theta_0=s_W^2+\delta
s_W^2=s_W^2-\frac{s_W^2c_W^2}{c_W^2-s_W^2}\left[\frac{\delta
G_F}{G_F}+\frac{\delta M_Z^2}{M_Z^2}\right] \nonumber \\
&=&
s_W^2+\frac{s_W^2c_W^2}{c_W^2-s_W^2}\left[\frac{\Delta}{4}\frac{(g^2-g^{\prime
2})^2}{g^4}+\frac{\Delta}{2} \left( \cot^2 \gamma \cos^2 \beta +
\tan^2 \eta \sin^2 \beta \right)\right],
\end{eqnarray}
where $\Delta=v^2/f^2$. We need to identify the photon and neutral
current couplings so as to calculate the shift in the couplings of
the fermions
\begin{eqnarray}
\mathcal{L}_{\mbox{nc}}&=& e A_{\mu} J^\mu_{Q} + \frac{e}{s_W
c_W}Z_\mu\left[ J^{3 \mu} \left(1 +
\frac{\Delta}{4}\frac{g^{\prime 2}(g^2-g^{\prime 2})}{g^4}\right)
\right. \nonumber
\\ && \left. - J^{\mu}_Q s_W^2\left(1 +\frac{\Delta}{4}\left(1-\frac{g^{\prime 4}}{g^4}\right)\right)+J^{15\mu}\frac{\Delta}{2\sqrt{2}}\left(1-\frac{g^{\prime 2}}{g^2}\right)\right]\nonumber \\
&& - \frac{g^{\prime 4}}{2 g^4
f^2}\left(J_3^{2}+J_Q^{2}\right)-\frac{1}{f^2}J_{15}^2+\frac{g^{\prime
4}}{f^2 g^4} J_3J_Q+\frac{\sqrt{2} g^{\prime^2}}{g^2 f^2}
J_{15}^{\mu}\left(J_{Q\mu}-J_{3\mu}\right)~ . \label{neutraldude}
\end{eqnarray}

In addition, we must also calculate the effects of fermion mixing
in this model, as $SU(2)$ doublet quarks mix with the heavy
singlet quarks.  This results in an additional shift of the
couplings of the doublet quark mass eigenstates.  The Yukawa
couplings leading to the fermion mass matrix are given by
\begin{equation}
\mathcal{L}_{\mathrm{quarks}} = \left(\lambda_1 u_1^c
\Phi_1^\dagger +\lambda_2 u_2^c \Phi_2^\dagger +\lambda_3 u_3^c
\Psi_1^\dagger\right) Q+\frac{\lambda_d}{\Lambda^2} d^c \Phi_1
\Psi_1 \Psi_2 Q + \mathrm{h.c.},
\end{equation}
where $Q=(q,\chi_1,\chi_2)^T$.  This gives a fermion mass matrix
\begin{equation}
M_f = \left( \begin{array}{cccc}
0 & 0 & 0 & 0 \nonumber \\
0 & 0 & 0 & 0 \nonumber \\
\lambda_2 \frac{v}{\sqrt{2}} \cos \beta \cos \gamma & 0 & \lambda_2
f_2
& 0 \nonumber \\
\lambda_3 \frac{v}{\sqrt{2}} \sin \beta \sin \eta & 0 & 0 &
\lambda_3
f_3 \nonumber \\
\end{array} \right)
\end{equation}
where we have taken a light quark limit, $\lambda_d,\lambda_1 \approx
0$, and ignored CKM mixing.
This matrix is diagonalized by a biunitary transformation, of
which the physically significant portion is acting on the $SU(4)$
multiplet, mixing the $SU(2)$ doublet quarks with the heavy
singlets. This mixing causes a shift in the couplings of the left
handed up quarks to the $SU(4)$ gauge bosons.  The current interactions can be
parametrized by
\begin{equation}
\label{eq:neut}
A^a_\mu J^{a \mu} = A^a_\mu \bar{Q} \gamma^\mu  T^a Q
\end{equation}
(no sum over $a$) where $T^a$ is the generator corresponding to $A_\mu^a$.  For instance, the $T_3$ matrix is given by
\begin{equation}
T^3=\left( \begin{array}{cccc} 1/2 & 0 & 0 & 0 \\ 0 & -1/2 & 0 & 0 \\ 0 & 0 & 0 & 0 \\ 0 & 0 & 0 & 0
    \end{array} \right).
\end{equation}
Equation~(\ref{eq:neut}) can be rewritten as
\begin{equation}
A^a_\mu \bar{Q}_M U T^a U^\dagger Q_M
\end{equation}
where $Q_M$ are the fermion mass eigenstates, and $U$ is the matrix
which rotates $Q$ into this mass eigenbasis.  The term involving only
the light fermions coming from the $(1,1)$ component of the matrix $U
D^a U^\dagger$ is the standard model coupling of the up-type quarks plus
a correction.  As an example, this procedure leads to a change in the
coupling of the up quarks to the $Z$-boson given by
\begin{equation}
\delta \tilde{g}^{u}_L = - \frac{\Delta}{4} \left[ \cot^2
  \gamma \cos^2 \beta + \tan^2 \eta \sin^2 \beta \right].
\end{equation}

The Yukawas for the leptons are similar:
\begin{equation}
\mathcal{L}_{\mathrm{leptons}} = \left( \lambda_1^\nu \chi_{\nu
1}^c
  \Phi_2^\dagger + \lambda_2^\nu \chi_{\nu 2}^c \Psi_1^\dagger\right)
  L + \frac{\lambda_e}{\Lambda^2} e^c \Phi_1 \Psi_1 \Psi_2 L
\end{equation}
where $L=(l,\chi_{\nu 1},\chi_{\nu 2})$.  Performing the same
procedure, one finds that the couplings of the neutrinos to the $Z$ are
shifted by exactly the same amount as the up-type quarks.
The couplings of the light fermions to the $t_{15}$ gauge bosons are modified in a
similar fashion, which would lead to shifts in electromagnetic
couplings, but because the charges for the heavy fermions are
identical to the up-quark charge, there is no shift.  There are also
corrections to $J_{15}$ four-fermion operators, but these are higher order
in $\Delta$.

The contributions from (\ref{neutraldude}) give
\begin{equation}
\delta \tilde g^{ff}=\frac{\Delta}{4} \left(1-\tan^2 \theta_W\right)
\left[\tan^2 \theta_W (t_3^f -q^f)+\sqrt{2} t_{15}^f \right]  
\end{equation}
where $t_{15}=-1/\sqrt{8}$ for the left-handed fermions and $0$
for the right-handed fermions.

There are also corrections to the charged current couplings from mass
mixing of the light neutrinos and heavy fermions that effectively lead
to an additional shift in $G_F$.
This shift is given by
\begin{equation}
\delta G_F = -\frac{1}{\sqrt{2} v^2} \frac{\Delta}{2} \left[ \cot^2
  \gamma \cos^2 \beta + \tan^2 \eta \sin^2 \beta \right]
\end{equation}
that was included in the expression for $G_F$ in Eq.~(\ref{eqar:mart})
and is the origin of the angle-dependent correction to $s_0^2$ in 
Eq.~(\ref{eqar:mart}).

\subsection{Four-fermion operators}

There are additional four-fermion operators present in this model
that were not present in other models due to the presence of the
$J^{15}$ currents.  To calculate the contributions to the
low-energy observables such as atomic parity violation we use a
slightly different method than used in~\cite{bigcorrections}. We
first integrate out the Z boson and obtain our low energy neutral
current Lagrangian
\begin{eqnarray}\label{lowneutraldude}
\mathcal{L}_{nc}&=&-\frac{g^2+g'^2}{2 M_Z^2}\left[ J^{3 \mu}
\left(1 + \frac{\Delta}{4}\frac{g^{\prime 2}(g^2-g^{\prime
2})}{g^4}\right) - J^{\mu}_Q s_W^2\left(1
+\frac{\Delta}{4}\left(1-\frac{g^{\prime
4}}{g^4}\right)\right)\right.
\nonumber\\
&& \left.{} +J^{15\mu}\frac{\Delta}{2\sqrt{2}}\left(1-\frac{g^{\prime 2}}{g^2}
\right)\right]^2 -\frac{g^{\prime 4}}{2 g^4 f^2}
\left(J_3^{2}+J_Q^{2}\right)-\frac{1}{f^2}J_{15}^2\nonumber\\
&&{} +\frac{g^{\prime 4}}{f^2 g^4} J_3J_Q+\frac{\sqrt{2}
g^{\prime^2}}{g^2 f^2} J_{15}^{\mu}\left(J_{Q\mu}-J_{3\mu}\right),
\end{eqnarray}
we then put relevant parts of the Lagrangian in the form
\begin{equation}
\mathcal{L}^{\nu\rm{Hadron}}=-\frac{G_F}{\sqrt{2}}\,\bar{\nu}\gamma^\mu(1-\gamma^5)\nu\times\sum_i\left[\epsilon_L(i)
\bar{q}_i
\gamma^\mu(1-\gamma^5)q_i+\epsilon_R(i)\bar{q}_i\gamma^\mu(1+\gamma^5)q_i\right],
\end{equation}

\begin{equation}
\mathcal{L}^{\nu\rm{e}}=-\frac{G_F}{\sqrt{2}}\,\bar{\nu}\gamma^\mu(1-\gamma^5)\nu\,\bar{e}\gamma_\mu(g_V^{\nu
e}-g_A^{\nu e}\gamma^5)e,
\end{equation}
and
\begin{equation}
\mathcal{L}^{e\rm{Hadron}}=\frac{G_F}{\sqrt{2}}\,\times\sum_i\left[C_{1i}\bar{e}\gamma_\mu\gamma^5
e\bar{q}_i\gamma^\mu q_i+C_{2 i}\bar{e}\gamma_\mu
e\bar{q}_i\gamma^\mu\gamma^5 q_i\right].
\end{equation}
We then use the expressions in~\cite{pdg} to calculate the low
energy observables for instance the ``weak'' charge of heavy atoms
is given by
\begin{equation}
Q_W=-2\left[C_{1 u}(2 Z+N)+C_{1 d}(Z+2 N)\right]
\end{equation}
where in this model
\begin{eqnarray}
C_{1u}=&&-\frac{1}{2} + \frac{4}{3} s_W^2 + \frac{2 \Delta}{3} s_W^2
\left[ 2 + \cot^2 \gamma \cos^2 \beta+\tan^2 \eta \sin^2 \beta \right] \nonumber \\
C_{1d}=&&\left(\frac{1}{2}-\frac{2}{3} s_W^2 \right)
  \left[1+\frac{\Delta}{2} \left\{2 + \cot^2 \gamma \cos^2 \beta +
  \tan^2 \eta \sin^2 \beta \right\} \right]
\end{eqnarray}
once the shift in the couplings due to the fermion mixing has also
been taking into account.

The tree level expressions for the low energy observables are then
\begin{eqnarray}
g_L^2 (\nu N \rightarrow \nu X) &=& \frac{1}{2} - s_W^2 + \frac{5}{9}
s_W^4 +\frac{\Delta}{12} (4 s_W^2 -3) \left[ \cot^2 \gamma \cos^2
  \beta + \tan^2 \eta \sin^2 \beta \right] \nonumber \\
g_R^2 (\nu N \rightarrow \nu X) &=& \frac{4}{9} s_W^4 \nonumber \\
g_{eA} (\nu e \rightarrow \nu e) &=& -\frac{1}{2} \nonumber \\
g_{eV} (\nu e \rightarrow \nu e) &=& -\frac{1}{2} +2 s_W^2 \nonumber \\
Q_W (\mathrm{Cs}) &=& -376 C_{1u} - 422 C_{1d},
\end{eqnarray}
where it is important to note that $s_W^2$ must be expressed in terms
of the observable $s_0^2$ in (\ref{eqar:mart}) to obtain the total shift of each parameter.

Direct experimental constraints on other four-fermion operators 
can also give bounds to little Higgs models.  Operators such as
\begin{equation}
\frac{1}{f^2} \bar{e}_L\gamma_\mu e_L \bar{q}_L\gamma^\mu q_L
\end{equation}
coming from the left-left currents in (\ref{lowneutraldude}) can contribute to
scattering processes that
have stringent bounds which can be found in~\cite{kingman}. The
bounds in this particular model are weaker than those coming from
the global fit of precision EW data so we will not review them
further.  Note that these operators arise in all little
Higgs models (to date), however we find the bounds from precision
electroweak data are stronger than the bounds on these four-fermion
operators.

\subsection{Numerical results}

We compute the bounds on this model using the bare expressions
re-expressed in terms of the physical input parameters as done in
the previous sections and in~\cite{bigcorrections}.  In the
general case of $f_{12}\neq f_{34}$ there would be additional
mixing between the $Z$, the $Z^\prime$, and  the gauge boson
corresponding to $T^{12}$. One could take into account the angle
describing the ratio $f_{12}/f_{34}$ and do another fit, however
this adds considerable complexity to the analysis due to mass
mixing of heavy gauge eigenstates, and we leave such a study for
future exploration.

We compute the bounds for a $95\%$ C.L. which for a $4$ parameter
fit corresponds to a $\Delta \chi^2=9.49$.  The minimum bound on
this model at a $95\%$ C.L. is $f = f_{12}=f_{34}=4.2$~TeV, which
occurs at $\tan \beta =0.1$, $f_2/f_1=10$, and $f_4/f_3=0.1$. Note
that we have constrained the ratios of all VEVs to be between
0.1 and 10.
\begin{figure}[h!t]
\centerline{\includegraphics[width=.9\hsize]{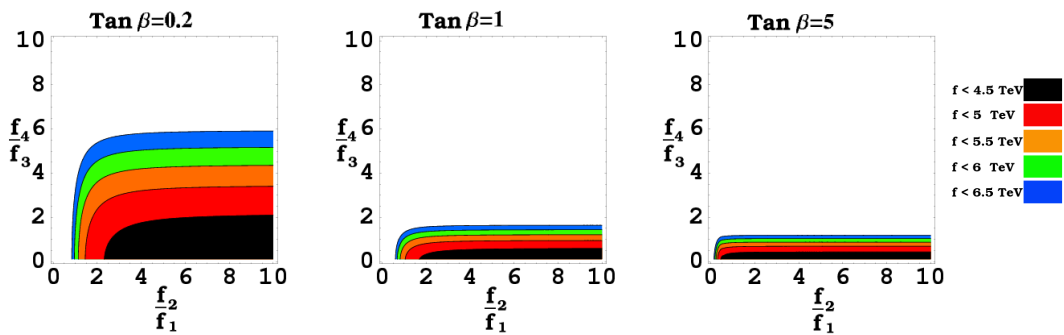}}
\caption{Plot of 95\% confidence level values of
$f=\sqrt{f_1^2+f_2^2}
  = \sqrt{f_3^2+f_4^2}$ as a function of $\tan \beta$, $f_2/f_1$, and
  $f_3/f_4$.} \label{martinplt}
\end{figure}
The minimum $\chi^2$ per degree of freedom for this model is
\begin{equation}
\frac{\chi^2}{\mbox{d.o.f.}}=1.67
\end{equation}
which we find is worse than the SM with a
$115$~GeV Higgs mass, which has $\chi^2/\mathrm{d.o.f.}=1.38$.
Raising the Higgs mass to $300$~GeV
raises the goodness of fit parameter to $\chi^2_{300}/\mathrm{d.o.f.} =
1.74$ with $m_H = 300$~GeV, for which
$\chi^{2\mathrm{SM}}_{300}/\mathrm{d.o.f.} = 1.59$.  Overall, the $115$~GeV Higgs
is still preferred.

\begin{figure}[h!t]
\centerline{\includegraphics[width=1\hsize]{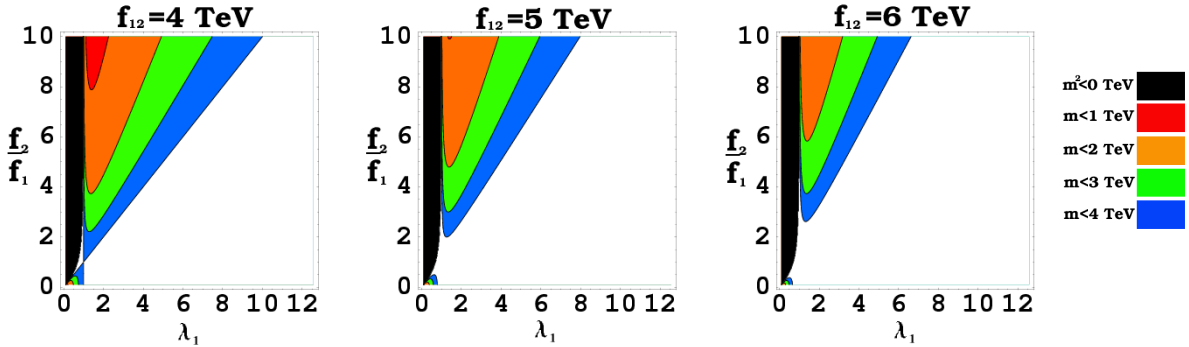}}
\caption{Plot of $m_\chi$ as a function of $\lambda_1$ and
$f_2/f_1$, with $\tan \beta = 0$.
  The black region corresponds to regions of parameter space that are
  forbidden due to a negative mass squared for the heavy fermion.
} \label{fig:FTplotmart}
\end{figure}

We would like to develop some way of interpreting the bounds on
$f$ in terms of a quantification of fine tuning.  In order to do
this, we calculate the mass of the heavy fermion as a function of
the ratio $f_2/f_1$ and the Yukawa coupling $\lambda_1$, fixing
$f$ at different values, and setting the top Yukawa $\lambda_t =
1$ to eliminate $\lambda_2$.  We choose the best case scenario, $\tan
\beta \approx 0$, where there is a larger region of $f$ below
$4.5$~TeV.  The results are shown in
Fig.~\ref{fig:FTplotmart}.  Therefore there exist regions
where $m_\chi<f$ and a large $f$ does not directly imply large fine
tuning.

\section{Conclusions}
\label{conclusions-sec}\setcounter{equation}{0}

We have calculated the tree-level bound on the symmetry breaking scale $f$
in several modifications of the littlest Higgs model. Some of these
modifications are motivated by trying to avoid large contributions to
EW precision observables. We find
that generically $f$ must be larger than several TeV\@.
Non-generically, however, we find the bound on $f$ can be relaxed
from $4$ TeV \cite{bigcorrections} to $1$-$2$ TeV depending on the model
variation and the degree of tuning of model parameters.
For the $SU(6)/Sp(6)$ model
we find that the minimal model has slightly lower bounds than the
littlest Higgs, $f > 3$ TeV\@.
Interestingly, we showed that a variation of the $SU(6)/Sp(6)$ model
exists with a larger region of parameter space where the bound on $f$
is near $1$~TeV\@.  The reason why the $SU(6)/Sp(6)$ can be modified
more
successfully is because it does not have a Higgs triplet, and because
the heavy $U(1)$ gauge boson is somewhat heavier than in the littlest Higgs.  The triplet
VEV in the littlest Higgs model generically reduces the size of the
allowed parameter space even in the case of gauging just $U(1)_Y$.
The $SU(4)^4/SU(3)^4$ model is bounded by
$f^2\equiv f_1^2+f_2^2>(4.2\ \mathrm{TeV})^2$\@.  In certain models (for
example $SU(4)^4/SU(3)^4$) a large $f$ does not directly
imply a large amount of fine tuning since the heavy fermion masses
that contribute to the Higgs mass can be lowered below $f$ for a
carefully chosen set of parameters.

In general there can be three sources of custodial $SU(2)$ violation
in little Higgs models: when light fermions
couple to the heavy $U(1)$ gauge boson, when there is an $SU(2)$ triplet
that acquires a VEV, or when there are two Higgs doublets with unequal
expectation values ($\tan\beta \not= 1$).  It is interesting that
the Littlest Higgs model and the $SU(6)/Sp(6)$ each have a source
of custodial $SU(2)$ violation arising from the larger scalar sector
of the model with a separate parameter ($a$ in the Littlest Higgs model,
$\tan\beta$ in $SU(6)/Sp(6)$) characterizing the size of this
contribution.  There has also been a interesting
recent proposal for a little Higgs model that incorporates an approximate
custodial $SU(2)$ symmetry \cite{wackchang}, based on a variation
of the original minimal moose model \cite{minmoose}.  This model is
constructed specifically to avoid the gauge contributions to
custodial $SU(2)$ violation, although a scalar $SU(2)$ triplet is present
which may get an expectation value.  We will report on the EW constraints
on this model elsewhere \cite{future}.

The $SU(4)^4/SU(3)^4$ model has a qualitatively new feature with
respect to EW precision constraints.  In this model
the quadratically  divergent contributions of the Higgs are canceled by
gauge bosons mostly in $SU(4)$ that are themselves orthogonal.  The
absence of mixing of light with heavy gauge bosons is a desirable
feature to avoid EW constraints.  This is exact for the $SU(2)$
gauge bosons, but only approximate for the $U(1)$ gauge bosons 
after $SU(4)$ symmetry breaking.  In particular, given $f_{12} = f_{34}$, 
we found that EW precision observables were shifted 
by an amount proportional to $(g^2-g'^2)/g^2$, which characterizes the 
mixing of the additional $U(1)$ gauge boson with neutral $SU(4)$ gauge 
bosons.  If it had been possible to embed the $U(1)$ into some larger group
then with suitably arranged symmetry breaking it is likely that 
light/heavy gauge boson mixing could be eliminated.
(There would still be fermion mixing corrections.)  However, to obtain
\emph{both} the proper hypercharges for quarks and leptons as
well as the correct weak mixing angle, an additional $U(1)$ was required,
and thus mixing between light with heavy $U(1)$ gauge bosons was
inevitable.  It would be extremely interesting to determine if this
can or cannot be avoided in variations of this model.

Finally, several models contain two Higgs doublets that each acquire
expectation values.  It is well known that if each Higgs couples to
both up-type and down-type fermions together (unlike in the minimal
supersymmetric standard model, for example), there are additional
contributions to
flavor changing neutral current processes.  Since one motivation of
a larger cutoff scale in little Higgs models is to ensure that new
four-fermion flavor-violating operators are not strongly constraining,
this possible new source of flavor violation must be curtailed.
We have already emphasized that a
UV completion should determine the viability of the small regions
of parameter space where $f$ can be lowered into the $1$-$2$ TeV
range.  For models with two Higgs doublets, the UV completion should
also determine the fermion couplings to the Higgs multiplets.
For example, a supersymmetric completion may enforce holomorphic
superpotential couplings, thereby avoiding this source of FCNC.
It would be very interesting to see what restrictions the absence of
new FCNC violating Higgs couplings imposes on the UV completion.

%%%%%%%%%%%%%%%%%%%%%%%%%%%%%%%%%%%%%%%%%%%%%%%%%%%%%%
\section*{Acknowledgments}

We thank Nima Arkani-Hamed, Kingman Cheung, Andy Cohen, Thomas
Gregoire, Martin Schmaltz, and Jay Wacker for helpful discussions.  We
also thank Martin Schmaltz for useful comments on the first version of
this paper.  G.D.K. thanks the theory groups at LANL and CERN for hospitality
where part of this work was done.  The research of C.C., J.H., and
P.M. is supported in part by the NSF under grant PHY-0139738, and
in part by the DOE OJI grant DE-FG02-01ER41206. The research of
G.D.K. is supported by the US Department of Energy under contract
DE-FG02-95ER40896. The research of J.T. is supported by the US
Department of Energy under contract W-7405-ENG-36.


\begin{thebibliography}{99}

\bibitem{LEPEWG}
LEP Electroweak Working Group, LEPEWWG/2002-01, \\
http://lepewwg.web.cern.ch/LEPEWWG/stanmod/.

\bibitem{little1}
N.~Arkani-Hamed, A.~G.~Cohen and H.~Georgi,
%``Electroweak symmetry breaking from dimensional deconstruction,''
Phys.\ Lett.\ B {\bf 513}, 232 (2001) {\tt [hep-ph/0105239]}.
%%CITATION = HEP-PH 0105239;%%

\bibitem{other1}
N.~Arkani-Hamed, A.~G.~Cohen, T.~Gregoire and J.~G.~Wacker,
%``Phenomenology of electroweak symmetry breaking from theory space,''
JHEP {\bf 0208}, 020 (2002) {\tt [hep-ph/0202089]}.
%%CITATION = HEP-PH 0202089;%%


\bibitem{minmoose}
N.~Arkani-Hamed, A.~G.~Cohen, E.~Katz, A.~E.~Nelson, T.~Gregoire
and J.~G.~Wacker,
%``The minimal moose for a little Higgs,''
JHEP {\bf 0208}, 021 (2002) {\tt [hep-ph/0206020]}.
%%CITATION = HEP-PH 0206020;%%

\bibitem{littlest}
N.~Arkani-Hamed, A.~G.~Cohen, E.~Katz and A.~E.~Nelson,
%``The littlest Higgs,''
JHEP {\bf 0207}, 034 (2002) {\tt [hep-ph/0206021]}.
%%CITATION = HEP-PH 0206021;%%

\bibitem{other2}
T.~Gregoire and J.~G.~Wacker,
%``Mooses, topology and Higgs,''
JHEP {\bf 0208}, 019 (2002) {\tt [hep-ph/0206023]}.
%%CITATION = HEP-PH 0206023;%%


\bibitem{su6sp6}
I.~Low, W.~Skiba and D.~Smith,
%``Little Higgses from an antisymmetric condensate,''
Phys.\ Rev.\ D {\bf 66}, 072001 (2002) {\tt [hep-ph/0207243]}.
%%CITATION = HEP-PH 0207243;%%

\bibitem{dudes}
D.~E.~Kaplan and M.~Schmaltz,
%``The Little Higgs from a Simple Group,''
{\tt hep-ph/0302049}.
%%CITATION = HEP-PH 0302049;%%

\bibitem{wackchang}
S.~Chang and J.~G.~Wacker,
%``Little Higgs and custodial SU(2),''
{\tt hep-ph/0303001}.
%%CITATION = HEP-PH 0303001;%%


\bibitem{bigcorrections}
C.~Cs\'aki, J.~Hubisz, G.~D.~Kribs, P.~Meade and J.~Terning,
%``Big corrections from a little Higgs,''
{\tt hep-ph/0211124}.
%%CITATION = HEP-PH 0211124;%%

\bibitem{hewett}
J.~L.~Hewett, F.~J.~Petriello and T.~G.~Rizzo,
%``Constraining the littlest Higgs,''
{\tt hep-ph/0211218}.
%%CITATION = HEP-PH 0211218;%%

\bibitem{burdman}
G.~Burdman, M.~Perelstein and A.~Pierce,
%``Collider tests of the little Higgs model,''
{\tt hep-ph/0212228}.
%%CITATION = HEP-PH 0212228;%%


\bibitem{wisconsin}
T.~Han, H.~E.~Logan, B.~McElrath and L.~T.~Wang,
%``Phenomenology of the little Higgs model,''
{\tt hep-ph/0301040}.
%%CITATION = HEP-PH 0301040;%%

\bibitem{Dib:2003zj}
C.~Dib, R.~Rosenfeld and A.~Zerwekh,
%``Higgs production and decay in the little Higgs model,''
{\tt hep-ph/0302068}.
%%CITATION = HEP-PH 0302068;%%

\bibitem{wisc2}
T.~Han, H.~E.~Logan, B.~McElrath and L.~T.~Wang,
%``Loop induced decays of the little Higgs: H $\to$ g g, gamma gamma,''
{\tt hep-ph/0302188}.
%%CITATION = HEP-PH 0302188;%%


\bibitem{Sekhar}
R.~S.~Chivukula, N.~Evans and E.~H.~Simmons,
%``Flavor physics and fine-tuning in theory space,''
Phys.\ Rev.\ D {\bf 66}, 035008 (2002)
{\tt [hep-ph/0204193]}.
%%CITATION = HEP-PH 0204193;%%


\bibitem{CST}
R.~S.~Chivukula, E.~H.~Simmons and J.~Terning,
%``Limits on the ununified standard model,''
Phys.\ Lett.\ B {\bf 346}, 284 (1995) {\tt [hep-ph/9412309]}.
%%CITATION = HEP-PH 9412309;%%

\bibitem{RSfit}
C.~Cs\'aki, J.~Erlich and J.~Terning,
%``The effective Lagrangian in the Randall-Sundrum model and electroweak  physics,''
Phys.\ Rev.\ D {\bf 66}, 064021 (2002), {\tt [hep-ph/0203034]}.
%%CITATION = HEP-PH 0203034;%%

\bibitem{CEKT}
C.~Cs\'aki, J.~Erlich, G.~D.~Kribs and J.~Terning,
%``Constraints on the SU(3) electroweak model,''
Phys.\ Rev.\ D {\bf 66}, 075008 (2002), {\tt [hep-ph/0204109]}.
%%CITATION = HEP-PH 0204109;%%

\bibitem{future}
C.~Cs\'aki, J.~Hubisz, G.~D.~Kribs, P.~Meade, and J.~Terning, to
appear.


\bibitem{Nima}
N.~Arkani-Hamed, private communication.



\bibitem{Sformulas}
M.~E.~Peskin and T.~Takeuchi,
%``Estimation of oblique electroweak corrections,''
Phys.\ Rev.\ D {\bf 46}, 381 (1992).
%%CITATION = PHRVA,D46,381;%%

\bibitem{ErlerLang}
J.~Erler and P.~Langacker, review in ``The Review of Particle
Properties,'' K.~Hagiwara {\it et al.}  [Particle Data Group
Collaboration],
%``Review Of Particle Physics,''
Phys.\ Rev.\ D {\bf 66}, 010001 (2002),
%%CITATION = PHRVA,D66,010001;%%
updated version online:
http://www-pdg.lbl.gov/2001/stanmodelrpp.ps.

\bibitem{Lynn}
D.~C.~Kennedy and B.~W.~Lynn,
%``Electroweak Radiative Corrections With An Effective Lagrangian: Four Fermion Processes,''
Nucl.\ Phys.\ B {\bf 322}, 1 (1989);
%%CITATION = NUPHA,B322,1;%%
D.~C.~Kennedy, B.~W.~Lynn, C.~J.~Im and R.~G.~Stuart,
%``Electroweak Cross-Sections And Asymmetries At The Z0,''
Nucl.\ Phys.\ B {\bf 321}, 83 (1989).
%%CITATION = NUPHA,B321,83;%%

\bibitem{Burgess}
C.~P.~Burgess, S.~Godfrey, H.~Konig, D.~London and I.~Maksymyk,
%``Model independent global constraints on new physics,''
Phys.\ Rev.\ D {\bf 49}, 6115 (1994) {\tt [hep-ph/9312291]}.
%%CITATION = HEP-PH 9312291;%%


\bibitem{GAPP}
J.~Erler,
%``Global fits to electroweak data using GAPP,''
{\tt hep-ph/0005084}.
%%CITATION = HEP-PH 0005084;%%

\bibitem{futuregregoire}
T.~Gregoire, D.~R.~Smith and J.~G.~Wacker,
%``What precision electroweak physics says about the SU(6)/Sp(6) little  Higgs,''
arXiv:hep-ph/0305275.
%%CITATION = HEP-PH 0305275;%%

\bibitem{doublettriplet}
K.~Inoue, A.~Kakuto and H.~Takano,
%``Higgs As (Pseudo)Goldstone Particles,''
Prog.\ Theor.\ Phys.\  {\bf 75}, 664 (1986);
%%CITATION = PTPKA,75,664;%%
A.~A.~Anselm and A.~A.~Johansen,
%``Susy GUT With Automatic Doublet - Triplet Hierarchy,''
Phys.\ Lett.\ B {\bf 200}, 331 (1988);
%%CITATION = PHLTA,B200,331;%%
Z.~G.~Berezhiani and G.~R.~Dvali,
%``Possible Solution Of The Hierarchy Problem In Supersymmetrical Grand Unification Theories,''
Bull.\ Lebedev Phys.\ Inst.\  {\bf 5}, 55 (1989) [Kratk.\
Soobshch.\ Fiz.\  {\bf 5}, 42 (1989)];
%%CITATION = SPLRD,5,55;%%
Z.~Berezhiani, C.~Cs\'aki and L.~Randall,
%``Could the supersymmetric Higgs particles naturally be pseudoGoldstone bosons?,''
Nucl.\ Phys.\ B {\bf 444}, 61 (1995) {\tt [hep-ph/9501336]}.
%%CITATION = HEP-PH 9501336;%%

\bibitem{pdg}
K.~Hagiwara {\it et al.}  [Particle Data Group Collaboration],
%``Review Of Particle Physics,''
Phys.\ Rev.\ D {\bf 66}, 010001 (2002).
%%CITATION = PHRVA,D66,010001;%%

\bibitem{kingman}
K.~M.~Cheung,
%``Constraints on electron quark contact interactions and implications to  models of leptoquarks and extra Z bosons,''
Phys.\ Lett.\ B {\bf 517}, 167 (2001), {\tt [hep-ph/0106251]}.
%%CITATION = HEP-PH 0106251;%%

\end{thebibliography}
\end{document}